\documentclass[twocolumn]{aastex62}

\submitjournal{ApJ}

\newcommand{\kms}{\ensuremath{}{\rm \, km~s^{-1}}}

\newcommand{\mbh}{\ensuremath{M_\mathrm{BH}}}

\usepackage{amsmath}

\shorttitle{Axisymmetry in Triaxial Schwarzschild Models}
\shortauthors{Quenneville et al.}

\begin{document}

\title{Dynamical Modeling of Galaxies and Supermassive Black Holes: Axisymmetry in Triaxial Schwarzschild Orbit Superposition Models}

\correspondingauthor{Matthew E. Quenneville}
\email{mquenneville@berkeley.edu}

\author{Matthew E. Quenneville}
\affiliation{Department of Astronomy, University of California, Berkeley, CA 94720, USA}
\affiliation{Department of Physics, University of California, Berkeley, CA 94720, USA}

\author{Christopher M. Liepold}
\affiliation{Department of Astronomy, University of California, Berkeley, CA 94720, USA}
\affiliation{Department of Physics, University of California, Berkeley, CA 94720, USA}

\author{Chung-Pei Ma}
\affiliation{Department of Astronomy, University of California, Berkeley, CA 94720, USA}
\affiliation{Department of Physics, University of California, Berkeley, CA 94720, USA}

\begin{abstract}
    We present a detailed analysis of the behavior of the triaxial Schwarzschild orbit superposition method near the axisymmetric limit.  Orbit superposition modeling is the primary method used to determine dynamical masses of supermassive black holes (\mbh) in nearby galaxies; however, prior studies have reported conflicting results when comparing the outcome from axisymmetric orbit codes with that from a triaxial orbit code in the axisymmetric limit. We show that in order to achieve (oblate) axisymmetry in a triaxial code, care needs to be taken to axisymmetrize the short-axis tube orbits and to exclude both the long-axis tube and box orbits from the orbit library.
    Using up to 12 Gauss-Hermite moments of the line-of-sight velocity distributions as constraints, we demonstrate the effects of orbit types on the best-fit \mbh\ in orbit modeling of the massive elliptical galaxy NGC~1453 reported in \citet{Liepoldetal2020}. In addition, we verify the efficacy of our updated code on a mock galaxy dataset. We identify a subset of slowly precessing quasi-planar orbits for which the typical integration times can be insufficient to fully capture the equilibrium orbital behavior in both axisymmetric and triaxial systems with central black holes. Further investigation is needed for a more reliable treatment of these orbits.
\end{abstract}


\section{Introduction}
\label{sec:introduction}

The orbit superposition method of \citet{Schwarzschild1979} enables efficient construction of self-consistent and equilibrium mass models of galaxies.  The basic procedure consists of two steps: integrating a representative set of orbits in a static triaxial gravitational potential, and finding weights for these orbits such that their superposition reproduces the assumed mass distribution.

The orbit superposition method has been extended to include kinematic information and used to determine mass distributions in real galaxies, starting with studies such as \citet{Pfenniger1984, RichstoneTremaine1984, RichstoneTremaine1985, Rixetal1997}. From the quality of the fit to both kinematic and photometric data, this method can be used to assess the relative likelihood of a range of mass models and to determine best-fit mass parameters such as \mbh, stellar mass-to-light ratios, galaxy shapes, and dark matter halo parameters. 

Due to the large number of orbits needed to sample the relevant phase space, the orbit superposition method is computationally expensive.  To reduce the number of orbits and the dimensions of the model parameter space, a few orbit-based numerical codes have been developed for axisymmetric systems (e.g., \citealt{Crettonetal1999, Gebhardtetal2000a, Thomasetal2004, Vallurietal2004,  Cappellarietal2006}). Many dynamical measurements of \mbh\ from stellar kinematics have been obtained using these axisymmetric orbit codes. 

Triaxiality allows for more general galaxy shapes and additional orbit types, but modeling orbits in triaxial potentials comes at the cost of increased complexity and computation time. 
\citet{vandenBoschetal2008} presented a triaxial orbit-based code capable of comparing directly to observations, using an orbital sampling scheme based on \citet{Schwarzschild1993}. \citet{vandeVenetal2008} performed recovery tests of this code for analytically tractable triaxial potentials (excluding central black holes).
Only a handful dynamical determinations of \mbh\ have been obtained using triaxial models from this code \citep{vandenBoschdeZeeuw2010, Walshetal2012, FeldmeierKrause2017}.  Several additional \mbh\ were determined using this code in the (nearly) axisymmetric limit \citep{Sethetal2014, Walshetal2015, Walshetal2016, Walshetal2017, Ahnetal2018}. This code has also been used to construct axisymmetric and triaxial galaxy models to determine stellar dynamics and dark matter distributions for a wide range of galaxies (e.g., \citealt{Zhuetal2018b, Pocietal2019, Jinetal2020}). \citet{VasilievValluri2020} recently presented a new triaxial orbit-based code using a different method for phase space sampling and orbit initialization; the method was tested on mock data but had not been applied to real data.

An important test of the orbit superposition codes is the ability to produce consistent results between an axisymmetric code and a triaxial code in the axisymmetric limit. We note that the code by \citet{vandenBoschetal2008} is written for triaxial potentials and ``is not capable of making a perfectly axisymmetric model'' \citep{vandenBoschdeZeeuw2010}. Studies that attempt to run it near axisymmetry and then compare with results from axisymmetric codes have reached conflicting conclusions. For instance, \citet{vandenBoschdeZeeuw2010} used their triaxial code to construct (nearly) axisymmetric models for M32 and NGC3379, and found the mass-to-light ratios and \mbh\ to be consistent with those from earlier studies using axisymmetric codes (\citealt{vanderMareletal1998, Josephetal2001, Verolmeetal2002} for M32; \citealt{Gebhardtetal2000a, Shapiroetal2006} for NGC~3379).  \citet{Ahnetal2018}, on the other hand, found a puzzling global $\chi^2$ minimum at $\mbh = 0$ while using this triaxial code to perform axisymmetric modeling of the ultracompact dwarf galaxy M59-UCD3. They found this minimum to be inconsistent with the best-fit non-zero \mbh\ 
from Jeans modeling and the axisymmetric orbit code of \citet{Cappellarietal2006}.

It is the purpose of our recent work \citep{Liepoldetal2020} and this paper to investigate how to modify the \citet{vandenBoschetal2008} code to enable it to handle properly both axisymmetric and triaxial systems. 
Since no galaxy in nature is likely to be exactly axisymmetric, it may appear that we are taking a step backwards in examining the axisymmetric limit of a triaxial code.  While our next goal is indeed to adopt the more realistic triaxial potentials, we believe that one critical test of a triaxial code is its behavior in the simpler, axisymmetric limit.  Such a study -- the main goal of this paper -- is a particularly important step in the quest for dynamical \mbh\ measurements in view of the facts that almost all existing \mbh\ measurements have been obtained assuming exact axisymmetry, and that the aforementioned recent comparison of axisymmetric and triaxial codes have led to unresolved conflicting results. 

In \citet{Liepoldetal2020}, we described a set of recipes and code changes for achieving axisymmetry.  We then performed proper axisymmetric orbit modeling using the revised code to obtain a new \mbh\ measurement for the massive elliptical galaxy NGC~1453, a fast rotator in the MASSIVE survey \citep{Maetal2014} well suited for axisymmetric orbit modeling. Similar to \citet{Ahnetal2018}, we had encountered difficulties in constraining \mbh\ in NGC~1453 when we used the original code with comparable settings.  Through extensive testing, we came to two main conclusions: (1) higher Gauss-Hermite moments (beyond the typically used $h_4$) of the line-of-sight velocity distributions (LOSVDs) are needed to fully constrain the orbital weights, and (2) the orbit libraries need to be modified to satisfy axisymmetry. The use of higher moments is described in detail in \citet{Liepoldetal2020}. Here, we focus on the construction of axisymmetric orbit libraries in a triaxial orbit code.

 In this paper, we provide a full discussion of the required steps to axisymmetrize the model and the various modifications that we have implemented to the triaxial code by \citet{vandenBoschetal2008}.  The code was never given a name; we will refer to it as the TriOS (``Triaxial Orbit Superposition'') code from this point on. In Section~\ref{sec:background}, we provide some background information about the implementation of the orbit superposition method in this code.  We focus on four topics that are pertinent to subsequent discussions: the three major orbit types in a triaxial potential (Section~\ref{sec:orbittypes}), orbit sampling and initialization (Section~\ref{sec:orbitinitialization}), orbit integration (Section~\ref{sec:orbit_integration}), and parameters used to quantify triaxial shapes (Section~\ref{sec:triaxiality}).

In Section~\ref{sec:axi}, we give an in-depth discussion of the three main ingredients for axisymmetry listed in Section~4.1 of \citet{Liepoldetal2020}: axisymmetrization of short-axis tube orbits (Section~\ref{sec:axi_shorttubes}), criteria for how to exclude long-axis tube orbits (Section~\ref{sec:axi_longtubes}), and exclusion of box orbits (Section~\ref{sec:axi_boxes}).

We have made additional improvements and corrections to the code (Section~\ref{sec:minor_changes}). 
We identify a subset of slowly precessing quasi-planar orbits that are misclassified and are ``mirrored" improperly in the orbit library
(Section~\ref{sec:classification}). We correct an issue with the zero point of the logarithmic potential for the dark matter halo that would otherwise render energy conservation checks ineffective in the code (Section~\ref{sec:loghalo}).  
We are able to speed up the total runtime of a mass model by a factor of 2 to 3 by a simple modification to how the point spread function convolution is implemented in the code (Section~\ref{sec:psf}). 
An improvement in setting the intrinsic mass grid used to constrain stellar density profiles is described in Section~\ref{sec:massgrid}.
Finally, we illustrate the effects of these changes in the case of NGC~1453 (Section~\ref{sec:1453}).

Three appendices are included as well. Appendix~\ref{sec:ap_longaxistubes} derives a simple analytic criterion for the existence of long-axis tube orbits within a model. Appendix~\ref{sec:ap_thinorbits} outlines a change in the thin orbit finding algorithm that must be made to the TriOS code in order to generate the correct orbit sampling. Finally, Appendix~\ref{sec:ap_mocktest} presents a mock recovery test demonstrating the ability of our revised TriOS code to recover the input mass parameters.

\section{Orbit Modeling Background}
\label{sec:background}

A summary of the implementation of the Schwarzschild orbit superposition method in the TriOS code is given in Section~4 of \citet{Liepoldetal2020}. Here we focus on the topics relevant for subsequent discussions of axisymmetry (Section~\ref{sec:axi}) and code modifications (Section~\ref{sec:minor_changes}).

In this paper, we use a Cartesian coordinate system in which the $x$, $y$, and $z$ axes are directed along the intrinsic major, intermediate, and minor axes of the galaxy, respectively.  The $z$-axis is therefore the symmetry axis of an oblate axisymmetric potential, and the $x$-axis is the symmetry axis of a prolate axisymmetric potential.
We focus on oblate axisymmetric systems in this paper, although our discussions can be easily modified for the prolate axisymmetric case.

\subsection{Orbit Types in a Triaxial Potential}
\label{sec:orbittypes}

In a static triaxial gravitational potential, time invariance is the only global continuous symmetry of the Hamiltonian, $H$. By Noether's theorem, this symmetry gives rise to conservation of energy as the only ``classical'' integral of motion. This conservation law restricts the allowed phase space for a given orbit from the full six phase space dimensions to a five dimensional subspace defined by the energy $H=E$.  An integral that reduces the allowed phase space dimension in this way is referred to as an isolating integral.

Numerical studies have revealed that orbits in many potentials often conserve two additional ``non-classical'' isolating integrals of motion \citep{Schwarzschild1979}, which we refer to as $I_2$ and $I_3$. These additional integrals do not typically have simple analytical expressions nor correspond to global symmetries of $H$. Orbits that conserve three (or more) isolating integrals of motion are referred to as regular. These regular orbits often fall into one of three main orbit types: short axis tubes, long axis tubes, and boxes. 

\begin{figure*}
  \centering
  \includegraphics[width=\linewidth]{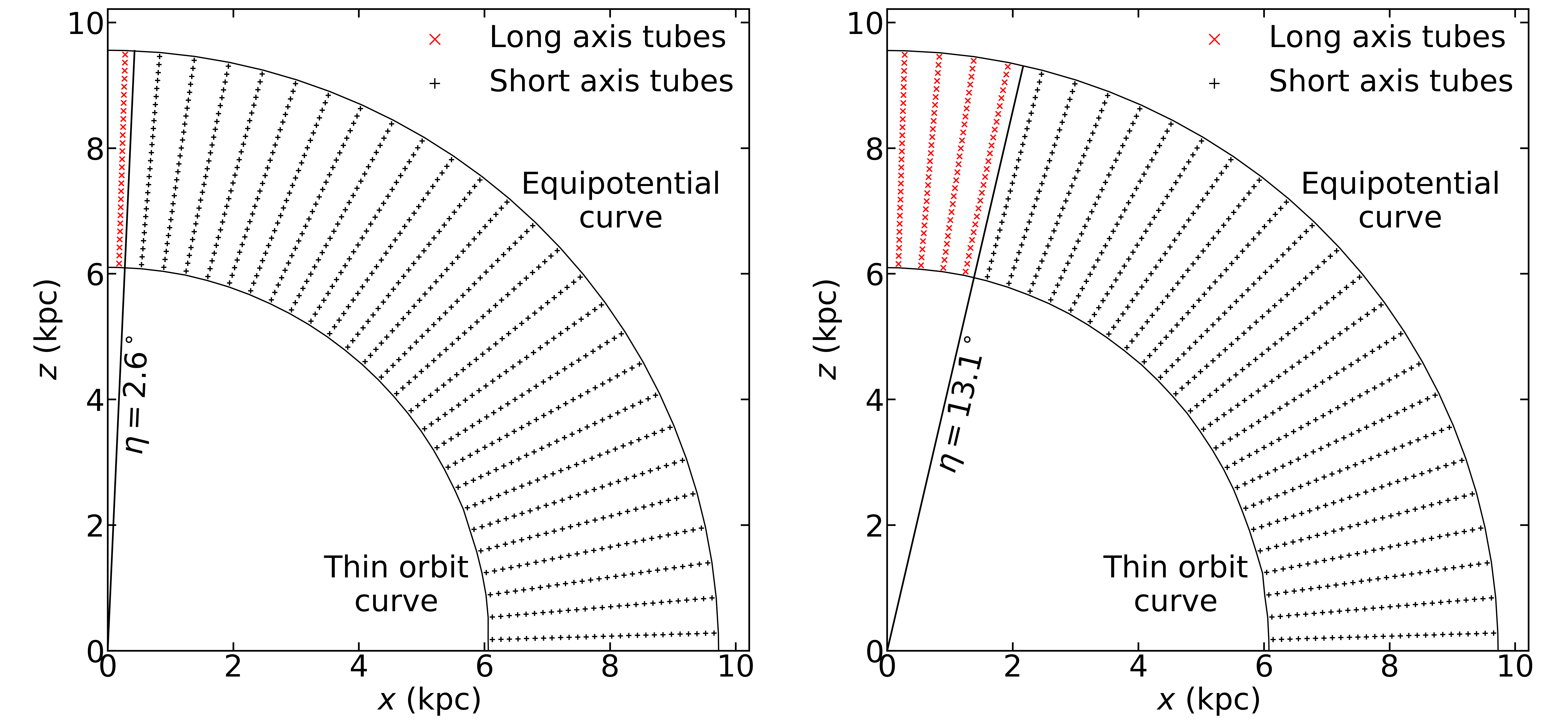}
  \caption{Two examples of the initial orbit locations in the $x$-$z$ start space.  Two nearly axisymmetric models for massive elliptical galaxy NGC~1453 are shown: (left) triaxiality parameter $T=0.002$, (luminosity weighted) axis ratio $p=0.9997$, and viewing angles $(\theta, \phi, \psi)=(89^\circ, 45^\circ, 90.001^\circ)$; (right) $T=0.05$, $p=0.993$, and $(\theta, \phi, \psi)=(89^\circ, 45^\circ, 90.026^\circ)$. Both models have the best-fit \mbh, mass-to-light ratio, and dark matter halo from \citet{Liepoldetal2020} and assume the orbit sampling parameters $(N_\Theta, N_R, N_\mathrm{Dither})=(9, 9, 3)$ (see Section~\ref{sec:axi_longtubes}).  In each panel, one energy is shown, where the energy is chosen such that the potential is dominated by the stellar mass. Each symbol represents the initial location for a single trajectory, which are bundled with adjacent trajectories to form one dithered orbit. The long-axis tubes (red crosses) are all contained within the angle $\eta$ of the $z$-axis for both values of $T$, where $\eta$ and $T$ are related by Equation~(\ref{eta}).  In general, more triaxial potentials contain a larger fraction of long-axis tubes in the $x$-$z$ start space.
  \label{figure:psi_startspace}}
\end{figure*}

Both types of tubes have a fixed sense of rotation. For short-axis tubes, the component of angular momentum along the potential's minor axis, $L_z$, does not change sign. Similarly, for long-axis tubes, the component of angular momentum along the potential's major axis, $L_x$, does not change sign. For box orbits, all three components of angular momentum change sign, leaving no fixed sense of rotation. Box orbits also have the property of touching the equipotential surface, $\Phi(x,y,z)=E$, at some point during their trajectory. Intermediate axis tube orbits are typically unstable in triaxial models \citep{HeiligmanSchwarzschild1979}. 

A triaxial system generally admits all three of these main orbit types. For oblate axisymmetric systems, the orbit structure is simpler because $L_z$ is an integral of motion, and only short-axis tubes are present.
Similarly, for prolate axisymmetric systems, $L_x$ is an integral of motion and only long-axis tubes are present.

\subsection{Orbit Sampling and Initialization}
\label{sec:orbitinitialization}

The set of initial conditions (referred to as a start space) should sample over all orbit types supported by the potential.
Even though regular orbits in a triaxial potential conserve energy plus two additional integrals of motion, 
the non-classical integrals of motion, $I_2$ and $I_3$, may not be the same quantities for each orbit type \citep{BinneySpergel1984,binneytremaine2008}. 
Thus, for a given energy, each orbit type can be sampled by a 2D start space, but the start spaces for the different orbit types cannot necessarily be combined into a single 2D start space.  

\citet{Schwarzschild1993} argued that a 4D space can guarantee that all orbit types of a given energy are sampled, and further suggested that a pair of 2D start spaces is sufficient for sampling phase space in realistic galaxy potentials.  The first of these start spaces, the $x$-$z$ start space, is defined by sampling over a grid of points in the $x$-$z$ plane, and setting $y=v_x=v_z=0$ and $v_y$ from $v_y^2= 2[E-\Phi(x,0,z)]$ for a given $E$. 
For simplicity, $v_y$ is taken to be positive and a second copy is added to the orbit library with the velocity direction flipped. 
Two examples of this $x$-$z$ start space are shown in Figure~\ref{figure:psi_startspace}.

Typically, tube orbits will pass through the positive quadrant of the $x$-$z$ plane perpendicularly at two points, separated by the thin orbit curve (see Figure~\ref{figure:psi_startspace}). Orbits launched along that curve will perpendicularly pass through the plane at a single point, so the curve can be found by iteratively launching orbits at different radii to identify those which pass through the $x$-$z$ plane in a thin curve (see Appendix~\ref{sec:ap_thinorbits}). Each orbit in the $x$-$z$ start space passes once inside and once outside the thin-orbit radius, so the code avoids double counting by initializing orbits only between the thin-orbit curve and the equipotential where $E = \Phi(r)$, as shown by the crosses in the examples in Figure~\ref{figure:psi_startspace}.
All three main orbit types pass through this start space. 

The second 2D start space proposed by \citet{Schwarzschild1993} is referred to as the stationary start space. In this start space, orbits are started from rest on the equipotential surface and are sampled over solid angle. Since tube orbits never come to rest, box orbits will be the only main orbit family in this start space. 
By combining the $x$-$z$ start space that samples mainly tube orbits with the stationary start space that samples mainly box orbits, \citet{Schwarzschild1993} suggests that any remaining unsampled region of phase space is likely to be small. 

The TriOS code is designed for static triaxial potentials that possess reflection symmetry along each of the three principal axes.  Under this assumption, any orbital property only needs to be calculated in one octant; it can then be ``mirrored'' into the other seven octants by symmetry.  Taking advantage of this symmetry, the code initializes orbits only in one octant ($x,y,z >0$) and integrates only these orbits. Seven additional copies of each orbit are then created by simply mirroring along the three axes. The details are described in Section~4.5 of \citet{vandenBoschetal2008} and the mirroring scheme is given in Table~2 there.  A key feature to note in Table~2 is that the exact mirror procedure (i.e., how the signs of the 
velocity components are flipped in each octant) depends on whether the orbit is a short-axis tube, long-axis tube, or box.
The orbits therefore must be classified first.

To classify an orbit, the code determines
how the angular momentum components change sign over the course of its integrated trajectory and uses these rules: (1) short-axis tubes, if $L_x$ and $L_y$ flip signs while $L_z$ does not, (2) long-axis tubes, if $L_y$ and $L_z$ flip signs while $L_x$ does not,
and (3) box orbits, if all three angular momentum components change signs.  The velocities are mirrored in order to maintain the orbit's sense of rotation. If an orbit does not fall into any of these categories, its velocity is mirrored to have zero angular momentum.

\subsection{Orbit Integration}
\label{sec:orbit_integration}
 
The TriOS code uses the \texttt{DOP853} explicit Runga-Kutta integrator with order 8(5,3). The integrator performs adaptive time stepping to ensure that the relative error in the positions and velocities are below a set threshold, typically $10^{-5}$.  After each orbit is integrated, a relative energy tolerance is used to check energy conservation. If the change in energy exceeds this tolerance (typically set to 10\%), it is re-integrated with a smaller integration error threshold.

The default integration time for each orbit is 200 dynamical times, where a dynamical time is set to the period of a closed elliptical orbit of the same energy.
To enforce smoothness of the recovered distribution function, the orbital initial conditions can be ``dithered'' by combining $N_{\rm Dither}^3$ trajectories corresponding to nearby initial conditions. By merging trajectories in this way, each orbit represents a small volume of the start space rather than a single point. This results in smoother orbital properties without a significant memory increase, since only the bundled orbital properties are stored.

After integration, the trajectory of each orbit is interpolated onto a set of points (typically 50,000) that are uniformly spaced in time.  These interpolated points are then stored and used for computing orbital properties.  Once the orbit libraries are constructed, weights are found for each orbit to reproduce the observed surface brightness (SB) distribution, the LOSVDs, and intrinsic 3D mass distribution. 

\subsection{Viewing Angles, Axis Ratios, and Triaxiality}
\label{sec:triaxiality}

Three viewing angles $(\theta, \phi, \psi)$ can be used to 
relate the intrinsic and projected coordinate systems of a triaxial galaxy \citep{Binney1985}. The two angles $\theta$ and $\phi$ describe the orientation of the observer's line of sight with respect to the intrinsic axes of the galaxy. The angle $\psi$ specifies the remaining degree of freedom -- rotation of the galaxy around the line of sight. The angle $\psi=90^\circ$ corresponds to an oblate axisymmetric potential.  In the oblate axisymmetric limit, $\theta$ is the inclination with $\theta=90^\circ$ corresponding to edge-on, and $\phi$ describes rotations about the symmetry axis.

These three viewing angles are related to the intrinsic axis ratios $p$ and $q$, where $p=b/a$ is the intrinsic intermediate-to-major axis ratio, $q=c/a$ is the intrinsic minor-to-major axis ratio, and $a, b, c$ are the lengths of the three principal axes of a triaxial system (with $c \le b \le a$). 
A third parameter, $u=a'/a$, represents a compression factor due to projection, where $a'$ is the major axis of the projected shape on the sky;
$u=1$ corresponds to the intrinsic major axis lying in the plane of the sky, while $u=p$ corresponds to the intrinsic intermediate axis lying in the plane of the sky. These quantities obey the inequality $0 \le q \le p \le u \le 1$.   The relationship between the viewing angles, intrinsic axis ratios and observed axis ratio is given in Equations~(7)-(10) of \citet{vandenBoschetal2008}. In addition, a triaxiality parameter is often used:
\begin{equation}
    T=\frac{1-p^2}{1-q^2} \,.
\label{T}
\end{equation}
This parameter ranges from 0 for oblate axisymmetry to 1 for prolate axisymmetry, with values in between indicating a triaxial shape.

The oblate axisymmetric limit can be achieved by setting either $p=1$ or $\psi=90^\circ$, but for numerical reasons, the code does not run when $\psi$ is exactly $90^\circ$. 
As we discussed in \citet{Liepoldetal2020} and elaborate below (Section~\ref{sec:axi_longtubes}), axisymmetry in the code can be achieved only with carefully chosen values of $\psi$ or $p$.

\section{Ingredients for Achieving Axisymmetry}
\label{sec:axi}

In this section, we discuss a number of steps that need to be taken to generate orbit-superposition models in the oblate axisymmetric limit using the TriOS code. It is straightforward to modify these steps for the prolate axisymmetric limit. In Appendix~\ref{sec:ap_mocktest}, we test the modified TriOS code on a mock dataset showing that we can accurately recover input parameters.

\subsection{Axisymmetrize Short-Axis Tube Orbits}
\label{sec:axi_shorttubes}

As we described in Section~\ref{sec:orbitinitialization}, a triaxial potential exhibits reflection symmetry along each principal axis, allowing the TriOS code to initialize orbits in only one octant of the $x$-$z$ start space. These orbits are then mirrored via eight-fold reflections about the principal axes into each of the other seven octants. 
This setup is not meant for axisymmetric systems,
in which the orbit library should respect azimuthal symmetry about the symmetry axis.

To enable modeling axisymmetric systems, we have implemented an axisymmetrized version of the orbit library by creating 80 copies of each short-axis tube orbit in the original loop library: 40 copies rotated evenly through an angle $2\pi$ about the short axis with velocities rotated to preserve $L_z$, and another 40 copies generated by flipping the sign of $z$ and $v_z$ in each of the 40 rotations. We choose 40 rotations, as this gives several copies per quadrant, with a comparable density to the start space grid sampling.
Once we perform this operation, it is unnecessary to perform the eight-fold reflections in the original code.  A similar rotation scheme was tested on mock data with no central SMBH in \citet{Hagenetal2019}.

The net result of our axisymmetrization process is to create a library of short-axis tube orbits in the TriOS code that samples the azimuthal angle uniformly with effectively equal orbital weights. In order for this procedure to be justified, the library should consist solely of short axis-tubes. In the next section, we show how to ensure that no long-axis tubes occur in this library.

\subsection{Exclude Long-Axis Tube Orbits}
\label{sec:axi_longtubes}

In an oblate axisymmetric potential, the long-axis tube orbits become unstable since there is no longer a single preferred long axis. These orbits therefore should not be present in the orbit library. \footnote{Similarly, in the case of a prolate axisymmetric potential, the short-axis tube orbits become unstable and should be absent.}

As we discussed in Section~\ref{sec:triaxiality}, the potential is oblate axisymmetric when $\psi$ is set to $90^\circ$ exactly, and long-axis tubes should be absent in this limit.  For numerical reasons, however, the code does not run when $\psi$ is set to $90^\circ$ within machine accuracy. Prior work using this code for black hole mass measurements in the axisymmetric limit chose either $\left|\psi-90^\circ\right|$ between $0.001^\circ$ and $0.01^\circ$~\citep{Walshetal2016,Ahnetal2018}, or an axis ratio of $p=0.99$~\citep{Sethetal2014,Walshetal2015,Walshetal2017}. As we first pointed out in \citet{Liepoldetal2020}, some of these values may not have been close enough to the desired axisymmetric values to exclude long-axis tubes. Here we provide a detailed explanation.

We use two examples of the $x$-$z$ start space in
Figure~\ref{figure:psi_startspace} to illustrate how long-axis tube orbits are initialized in the code. 

As shown in Appendix A, long-axis tube orbits in many realistic triaxial potentials are confined to pass through the $x$-$z$ start space within an angle $\eta$ from the $z$-axis. The angle $\eta$ depends on the shape of the potential, and we find the relation between $\eta$
and the triaxiality parameter $T$ (Equation~\ref{T}) to be well approximated by
\begin{equation}
     \eta=\tan^{-1}{\sqrt{\frac{T}{1-T}}} \,.
\label{eta}
\end{equation}
This is demonstrated in Figure~\ref{figure:psi_startspace} where the black line at angle $\eta$ separates the short-axis tube orbits (black crosses) from the long-axis tube orbits (red crosses).  As the potential becomes more oblate axisymmetric ($T=0.05$ in the right panel vs. $T=0.002$ in the left panel), $\eta$ becomes smaller and the area in the $x$-$z$ start space occupied by long-axis tubes shrinks. To effectively achieve oblate axisymmetry, $\eta$ needs to be small enough so that no orbits are sampled within an angle of $\eta$ of the positive $z$-axis.
Two additional mass models with higher triaxiality, ($T=0.25$ and 0.75) are shown in Appendix A and Figure~\ref{figure:triaxial_startspace}. Equation~(\ref{eta}) again provides an excellent approximation for the angle demarcating the long-axis and short-axis tube orbits in the $x$-$z$ start space.

Whether orbits are sampled within the angle $\eta$ on the $x$-$z$ plane depends on the input parameters.  For a given energy, the code starts the orbits on a grid of $N_R$ radii between the inner and outer thin orbit radii and $N_\Theta$ angles between $0^\circ$ and $90^\circ$ in the positive quadrant on the $x$-$z$ plane \citep{vandenBoschetal2008}. The code further allows for dithering, where $N_{\rm Dither}^3$ nearby initial conditions, adjacent in $(E,R,\Theta)$, are bundled together to improve the phase space sampling. Orbits are therefore sampled at a total of $N_\Theta\times N_{\rm Dither}$ angles, where the first angle from the $z$ axis is chosen to start at half of the grid spacing (i.e., at an angle of  $(\pi/2)/(2N_\Theta N_{\rm Dither})$ from the $z$-axis). The criterion to satisfy oblate axisymmetry is therefore 
\begin{equation}
    \label{bound}
    \frac{1}{2 N_\Theta N_{\rm Dither}} \frac{\pi}{2} \ga \eta.
\end{equation}

The two examples of NGC~1453 shown in Figure~\ref{figure:psi_startspace} have $N_\Theta=N_R=9$, $N_{\rm Dither}=3$, and $27\times 27$ orbits initialized in the $x$-$z$ start space.
The orbits closest to the $z$-axis are therefore at an angle of $\approx 1.67^\circ$ away.
These orbits lie within the demarcation angle $\eta$
of Equation~(\ref{eta}) for either model in Figure~\ref{figure:psi_startspace}:
$\eta = 2.56^\circ$ for $T=0.002$ (left) and
$\eta = 12.9^\circ$ for $T=0.05$ (right).
Both models therefore violate Equation~(\ref{bound}) and contain long-axis tubes.  This provides the physical explanation for our statement in \citet{Liepoldetal2020} that even $|\psi-90^\circ|$ as small as 0.001 (left panel) is not sufficiently close to $90^\circ$ to achieve axisymmetry in our models.

\begin{figure}
  \centering
  \includegraphics[width=\linewidth]{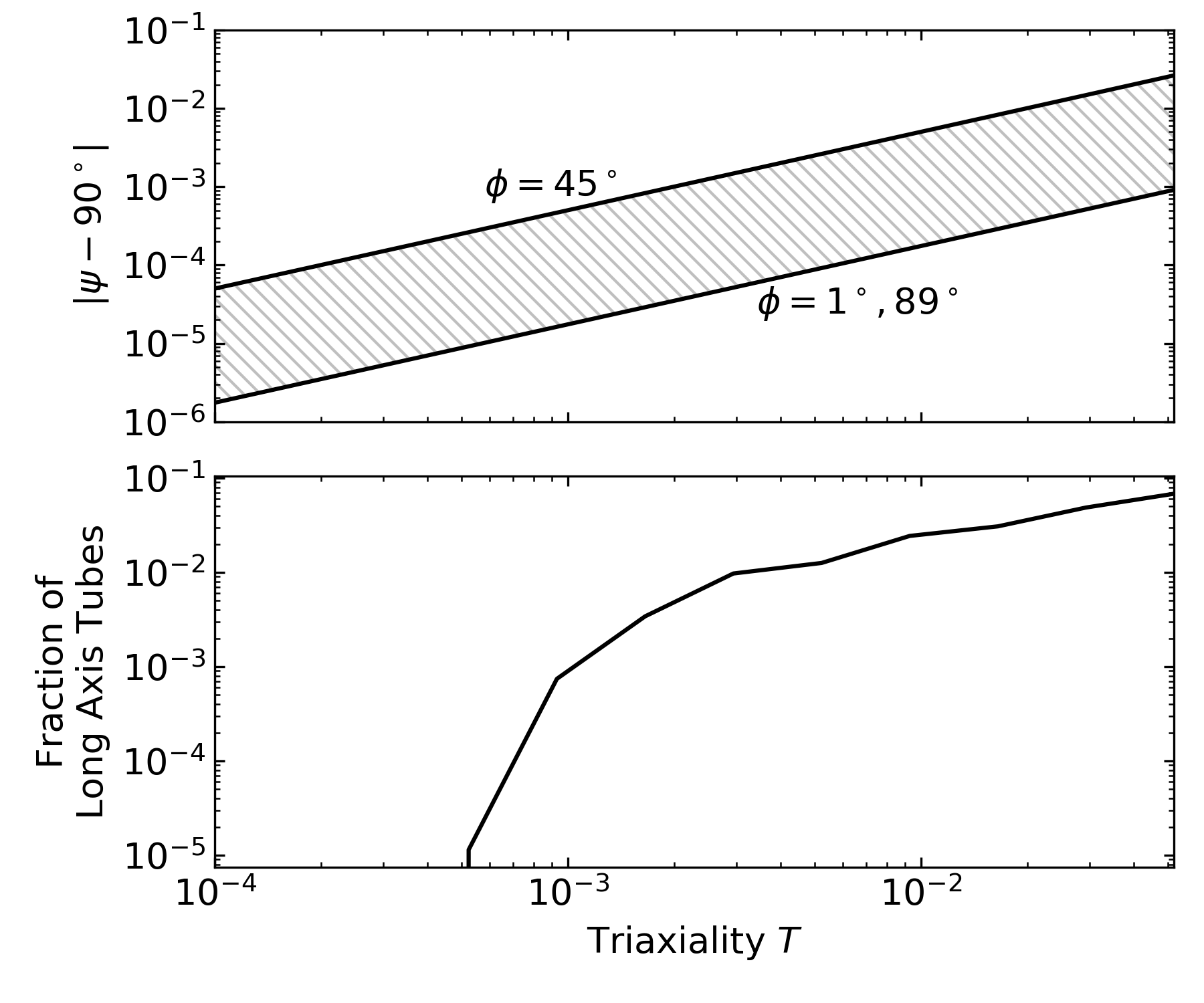}
  \hfill
\caption{(Top panel) Relationship between the viewing angle $\psi$ and the triaxiality of the deprojected stellar density. Exact oblate axisymmetry has $T=0$ and $\psi=90.0^\circ$. The other viewing angle $\theta$ is taken to be $89^\circ$, and $\phi$ is varied from $1^\circ$ to $89^\circ$. (Bottom panel) Fraction of long-axis tube orbits in the $x$-$z$ start space as a function of the triaxiality of the stellar density near the oblate axisymmetric limit. The same mass model and orbit sampling parameters for NGC~1453 shown in Figure~\ref{figure:psi_startspace} is assumed here. In this example, long-axis tube orbits begin to appear when $T$ is as small as $\sim 5\times 10^{-4}$ , or $|\psi-90^\circ|$ as small as $\sim 9\times 10^{-6}$,
and the fraction of these orbits increases monotonically as the potential becomes more triaxial, reaching $\sim 6\%$ at $T=0.05$.
\label{figure:psi}}
\end{figure}

To extend the discussion beyond the two specific mass models shown in Figure~\ref{figure:psi_startspace}, we illustrate in Figure~\ref{figure:psi} the relation between $T$ and $\psi$ for nearly axisymmetric models of NGC~1453 (top panel), and the corresponding fraction of long-axis tubes that are initialized in the $x$-$z$ plane (bottom panel). The inclination angle $\theta$ is assumed to be $89^\circ$ here, and the shaded band indicates the additional dependence of $T$ on $\phi$. 
Figure~\ref{figure:psi} shows that $T\la 5\times 10^{-4}$ is needed to exclude long-axis tube orbits in this case. The corresponding requirement on $\psi$ is $|\psi-90^\circ| \la 8.7\times 10^{-6}$ for $\phi\sim 1^\circ,89^\circ$ and $|\psi-90^\circ| \la 2.5\times 10^{-4}$ for $\phi\sim 45^\circ$.
We advocated $|\psi-90^\circ| = 10^{-9}$ in \citet{Liepoldetal2020}, which safely excluded all long-axis tube orbits. 

Earlier work using the code in the near axisymmetric limit does not typically satisfy the criterion in Equation~(\ref{bound}). For M59-UCD3, \citet{Ahnetal2018} used $(\theta, \phi, \psi) = (85^\circ, -49.99^\circ, 89.99^\circ)$, which we find to correspond to $T=0.004$ and $\eta = 3.64^\circ$.  The orbit sampling parameters were not explicitly given for the runs using the triaxial code.  Assuming the same parameters used in their runs with the axisymmetric orbit code ($N_\Theta = 8, N_{\rm Dither} = 6$), we find that the innermost ray would be at an angle of $0.94^\circ$ from the $z$-axis, which is well inside $\eta = 3.64^\circ$, and therefore violates the criterion in Equation~(\ref{bound}).

For M60-UCD1 \citep{Sethetal2014}, NGC~1271 \citep{Walshetal2015}, and Mrk~1216 \citep{Walshetal2017}, each paper quoted an axis ratio of $p=0.99$. The minimum possible triaxiality with this value of $p$ is $T=1-p^2=0.0199$ (in the unrealistic limit of a razor-thin disk with $q=0$), leading to a minimum $\eta$ of $8.1^\circ$.  For NGC~1271 and Mrk~1216, $N_{\rm Dither}=5$ was used, while $N_\Theta$ was set to 8 and 9 respectively. Thus, orbits were sampled starting at $1.125^\circ$ and $1^\circ$ away from the $z$-axis, indicating that neither satisfies the criterion in Equation~(\ref{bound}). For M60-UCD1, not enough information is given about the orbit sampling to determine whether the criterion is satisfied. However, for typical orbital sampling parameters quoted above, the criterion in Equation~(\ref{bound}) would not be satisfied.

The modeling of the NGC~1277 black hole used $N_\Theta=9$ and $N_{\rm Dither}=5$ \citep{Walshetal2016}; the innermost ray of initial orbits therefore lies at $1^\circ$ from the positive $z$-axis. The complete shape information was not given in the paper, but private communication indicated that $(\theta,\phi,\psi)=(75.3^\circ,71.6^\circ,90.001^\circ)$ was used. We find this set of viewing angles to correspond to $T=0.0002$ and $\eta = 0.85^\circ$, narrowly satisfying the criterion in Equation~(\ref{bound}).

We note that the presence of the long-axis tube orbits in the orbit library does not necessarily imply that they receive significant weights after fitting to observational constraints for a given galaxy.
Direct tests would need to be performed for each galaxy to assess the impact of these orbits on previous work.

\subsection{Exclude Box Orbits}
\label{sec:axi_boxes}

As we discussed in Section~\ref{sec:orbittypes}, all orbits in the (oblate) axisymmetric limit conserve $L_z$.  Box orbits in this limit have $L_z=0$ and therefore have similar properties as the tube orbits with small $L_z$.  In this case, as long as angular momentum is sufficiently sampled by the tube orbits, there is no need to include box orbits explicitly. 

The TriOS code devotes an entire library to box orbits and initializes them in the stationary start space (Section~\ref{sec:orbitinitialization}). One can modify the code to exclude this library when needed.  We use a simpler approach without changing the code itself: we skip running the orbit integration routine \texttt{orblib\_f.f90} for the stationary start space, and replace the box library with a copy of the $x$-$z$ library in the input file for the weight-finding routine \texttt{triaxnnls.f90}.  These modifications typically reduce the total computation time of the original code by more than half.

While box orbits are unnecessary in the axisymmetric limit, they also should be harmless and not affect the results if included. As a test, we have run our revised code including the box library for comparison.  Since the box orbits launched at different azimuthal angles are allowed to have different weights in the triaxial code, we have to impose an additional constraint of equal weights to enforce axisymmetry in the box library.
Once these weights are forced to be equal, we indeed find similar results as the case when the box library is excluded altogether. The case where the box library weights are free to differ between azimuthal angles is discussed in Section~\ref{sec:1453}. To reduce computational cost, we recommend excluding the stationary start space for axisymmetric models.

For a triaxial potential, we note that box orbits can also occur in the $x$-$z$ start space (e.g., Figure~1 of \citealt{Schwarzschild1993}). However, the region in the $x$-$z$ start space that would generate box orbits shrinks as the potential becomes increasingly axisymmetric. When exact axisymmetry is reached, only the orbits that begin exactly on the equipotential surface in the $x$-$z$ start space have $L_z=0$ (since they have zero initial velocities) and are box orbits. The TriOS code does not sample orbits lying exactly on the equipotential curve in the $x$-$z$ start space, so the number of box orbits will shrink to 0 as axisymmetry is approached.  In other orbit-based codes that assume axisymmetry from the start, the $L_z=0$ orbits also are not usually sampled, as they are presumed to be represented by the tube orbits with small but non-zero $L_z$ (e.g., \citealt{Crettonetal1999,Thomasetal2004}).

\section{Additional Code Fixes and Improvements}
\label{sec:minor_changes}

We have made several modifications in the TriOS code
in addition to those described in Section~\ref{sec:axi}. 
These modifications include corrections, improvements and speedups that are general to the code regardless of the issue of axisymmetry.  We describe these changes in this section.

\subsection{Correct Orbit Misclassifications}
\label{sec:classification}

As we described in Section~\ref{sec:orbitinitialization},
the TriOS code assumes the triaxial potential to possess reflection symmetry along each of the three principal axes and integrates only orbits that are initialized in one octant of space to save computation time. It then uses an eight-fold reflection scheme to generate seven more copies of each orbit. How the orbits are ``mirrored'' depends on whether the orbit is classified as a short-axis tube, long-axis tube, or box orbit.

We have discovered that the mirroring scheme in the original code misclassifies a subset of orbits for which the angular momenta vary on timescales slower than the integration time.  We find this to happen in at least two situations. First, in nearly oblate axisymmetric models, many box orbits in the stationary start space tend to be misclassified as short axis tubes due to the near conservation of $L_z$. Because $L_z$ varies slowly, it may not change sign throughout the integration time.  However, these orbits have very low angular momentum, so it is unlikely that mirroring these orbits to preserve $L_z$ would cause significant issues in the models themselves. 
 
The second situation occurs in regions of space where the potential is nearly spherical, e.g., deep within the SOI of an SMBH,
or in the outer part of a galaxy where a (spherical) dark matter dominates the potential. Some orbits in these regions follow quasi-planar rosettas or Keplerian-like ellipses with nearly constant angular momentum vectors, consistent with prior studies of orbits near a central point mass \citep{SridharTouma1999a,SridharTouma2016,Vallurietal2016}.
For the subset of orbits with precession time longer than the integration time, no component of their angular momentum changes sign over the entire integrated trajectories.
These orbits therefore do not fall into any of the categories listed above and are mirrored incorrectly to have no net angular momentum. 

These quasi-planar orbits will not be significant in most Schwarzschild models, as they are only present at extreme radii. We checked this in our models of NGC~1453, with the properly axisymmetrized code as described in Section~\ref{sec:axi} using the lowest four Gauss-Hermite kinematic moments as constraints. In this model, we find that $\sim 10\%$ of the total weight after orbital weight minimization is assigned to orbits that would have been quasi-planar in the original version of the code ($\sim 10\%$ of the mass within the Mitchell apertures and $\sim 2\%$ of the mass within the GMOS apertures). These relatively low percentages suggest a minimal effect on the model for NGC 1453.

We expect the issues with orbit integration time and misclassification to be more severe for galaxies with data that resolve well within the black hole's sphere of influence (SOI), or well beyond the stellar half-light radius, e.g., M87 and the Milky Way black hole. The effect is also likely to be more significant if the galaxy has a net rotation at these radii. 

We find a further issue with orbit classification in the orbital composition information outputted in the file \texttt{intrinsic\_moments.out}. This file reports the mass fraction of box orbits for each bin in the intrinsic spatial grid described in Section \ref{sec:massgrid}. In this case, however, all orbits that are neither long-axis tubes or short axis-tubes are grouped together as box orbits. Since this includes the quasi-planar orbits discussed above, the reported fraction of true box orbits may be overestimated.

In our revised code for axisymmetric systems, these orbit misclassification issues are not present because we manually assign all orbits as short-axis tubes and exclude all other orbit types. We will discuss further these quasi-planar orbits in triaxial systems in Section~\ref{sec:conclusion}.

\subsection{Fix Zero-point Issues with the Logarithmic Halo}
\label{sec:loghalo}

A logarithmic potential is often used to approximate the dark matter halo in prior orbit modeling work.  
The spherical version of a logarithmic halo is given by
\begin{equation}
    \Phi(r) = \frac{1}{2}V_c^2 \ln{\left(R_c^2+r^2\right)} + \Phi_0\,,
\label{eq:Phi}
\end{equation}
where $R_c$ is the core radius, and $V_c$ is the circular velocity at large $r$:
\begin{equation}
    V_c(r) = \frac{V_c\, r}{\sqrt{R_c^2 + r^2}}\,.
\label{eq:Vc}
\end{equation}

The zero point $\Phi_0$ can in principle be chosen arbitrarily; the original code set $\Phi_0=0$.
In practice, we find the choice of $\Phi_0=0$ and 
the use of physical units such as kilometers for all distances to create numerical problems.  The cause is simple: unlike other commonly used dark matter potentials such as \citet{Hernquist1990} and \citet{Navarroetal1996} that are negative at all locations and approach 0 at large $r$, the logarithmic potential with $\Phi_0=0$ is positive everywhere and grows unbounded at large $r$.
Thus, for the other potentials, $|\Phi(r)|$ can be interpreted as the local maximum kinetic energy for a bound orbit, but the orbital binding energy is infinite in the logarithmic potential. 
Furthermore, with the choice of $\Phi_0=0$, $|\Phi(r)|$ is much larger than the kinetic energy for all orbits in a logarithmic halo. This is because the central potential energy value, $\Phi(0) = V_c^2\ln(R_c)$, is much larger than the maximum possible kinetic energy sampled by the orbits, which is $\Phi(r_\mathrm{max})-\Phi(0)$,
where $r_\mathrm{max}$ is the largest equipotential radius of any orbit in a model. 

\begin{figure}
  \centering
  \includegraphics[width=\linewidth]{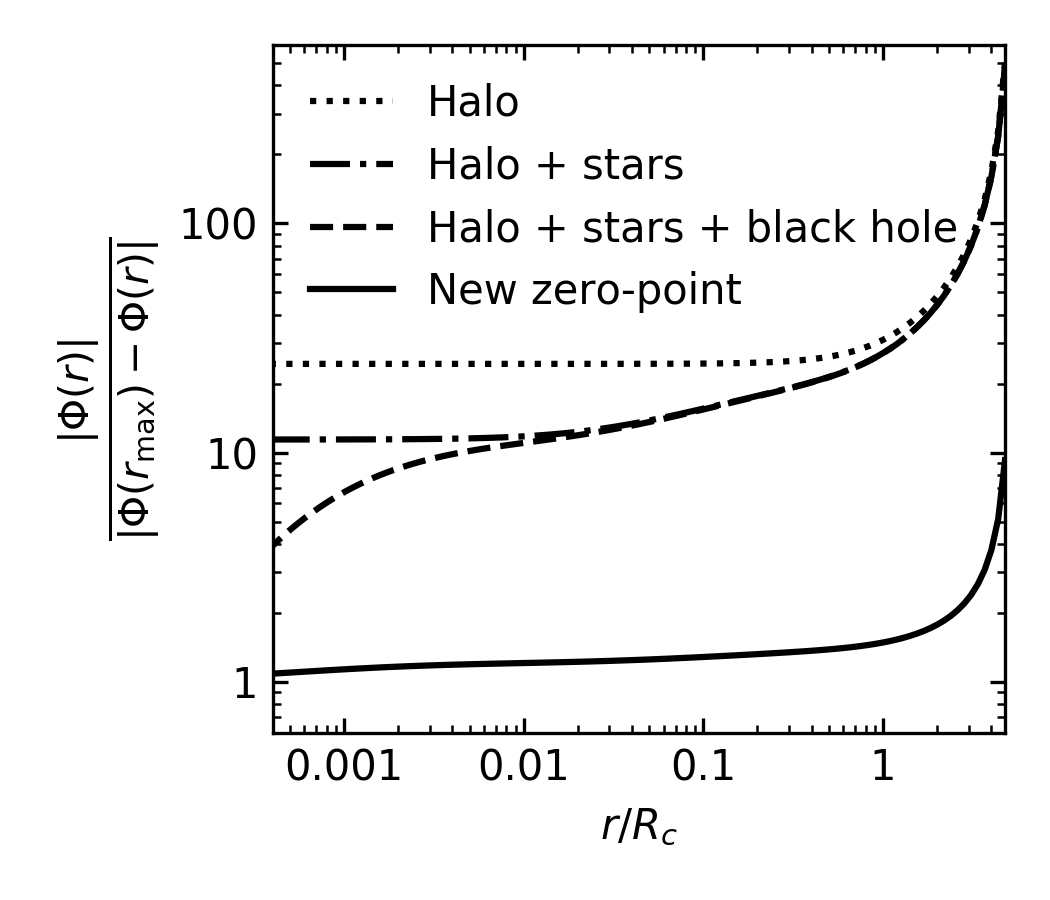}
  \hfill
  \caption{Illustration of the issue with setting the zero-point of the logarithmic potential to $\Phi_0=0$ in Equation~(\ref{eq:Phi}), as is assumed in the original code.  As an example, we use the best-fit mass model for   NGC 1453 in \citet{Liepoldetal2020} with a logarithmic dark matter halo of $R_c=15$ kpc and $V_c=633 \kms$.  The ratio of the potential energy to the maximum kinetic energy is plotted for this halo (dotted), halo plus stars (dot-dashed), and all three mass components (dashed). When this ratio is much larger than 1, as is shown for a large range of radius, even large errors in the kinetic energy would have little effect on the total energy. Energy conservation is therefore effectively not enforced in the original code for a logarithmic potential. The solid line shows the same ratio with all three mass components included, but with the halo zero point set according to Equation~\ref{eq:Phi0}.
  }
\label{figure:energy_conservation}
\end{figure}

To illustrate this point, we plot the ratio of $|\Phi(r)|$ and $|\Phi(r_\mathrm{max})-\Phi(0)|$ for the best-fit logarithmic dark matter halo of NGC~1453 \citep{Liepoldetal2020} in Figure~\ref{figure:energy_conservation} (dotted curve).
Additional contributions to the potential from the stars and black hole reduce the value of the potential energy and help lower this ratio (dot-dashed and dashed solid curves), but the ratio is well above unity for all relevant radii in all cases.

An unintended consequence of this large central offset is that even a $\sim 100\%$ change in the kinetic energy would contribute to only a tiny fraction of the total energy and would be difficult to detect.  The energy conservation checks in the code are therefore effectively not performed for most orbits.  While these numbers are worrying, we did not find the choice of $\Phi_0=0$ to affect significantly the best-fit mass parameters of NGC~1453 in \citet{Liepoldetal2020}. The reason for this particular case is that the orbit integrator happened to be accurate enough to satisfy the energy conservation tolerance (set to the default 10\%) even when this conservation criterion was unchecked.   
There is, however, no guarantee that this would be true for other galaxies or for parameters outside the ranges that we had explored.

To ensure energy conservation is checked in the code for the logarithmic potential, we choose a different zero point 
\begin{equation}
    \label{eq:Phi0}
    \Phi_0=-\Phi(r=2r_\mathrm{max}) \,,
\end{equation}
so that $\Phi(r)$ is negative for the entire allowed radial range of the orbits and approaches 0 outside the largest equipotential radius $r_\mathrm{max}$. The resulting ratio of $|\Phi(r)|$ to $|\Phi(r_\mathrm{max})-\Phi(0)|$ for the best-fit model of NGC~1453 is shown by the solid line in Figure~\ref{figure:energy_conservation}.

Our choice of $\Phi_0$ in Equation~(\ref{eq:Phi0}) also removes another issue that we have encountered with the original code: the orbit start space was sometimes not calculated correctly for mass models in which the black hole is either absent or has small mass compared to the stellar component and the logarithmic halo.
As discussed in Section~\ref{sec:orbitinitialization} and shown in Figure~\ref{figure:psi_startspace}, the $x$-$z$ start space of \citet{Schwarzschild1993} requires finding equipotential curves in the $x$-$z$ plane. The code locates it by finding the equipotential radius for each of a series of angles in the plane. For each angle, the equipotential radius is found via bisection with a relative tolerance that is typically taken between $10^{-7}$ and $10^{-5}$.  For $\Phi_0=0$, this tolerance again is not enforcing the intended accuracy level due to the large central value of $\Phi$.  For NGC~1453, this issue exists only for a few central equipotential radii and thus did not have a significant impact on our results.

\subsection{Speed Up Point Spread Function Implementation}
\label{sec:psf}

The point spread function (PSF) of the relevant observations needs to be incorporated into a mass model before the model is fitted to data to determine the orbital weights.
The TriOS code approximates the effect of the PSF by perturbing each trajectory at every stored time step with a pair of $\delta x$ and $\delta y$ randomly drawn from the PSF, which is assumed to be a single or multiple Gaussian functions.
This scheme involves a large number of operations since an orbit is typically stored at 50,000 points along the trajectory (see Section~\ref{sec:orbit_integration}),
and up to $\sim 10^6$ orbits can be used to represent a single mass model. 

The code generates each orbit perturbation by drawing two independent numbers, $k_x$ and $k_y$, from a uniform distribution over the interval $(-1,1)$ repeatedly until a pair with $k \equiv |\vec{k}| < 1$ is found. 
The perturbations $\delta x = \frac{k_x}{k} \sqrt{-2 \ln(k^2)}$ and $\delta y = \frac{k_y}{k} \sqrt{-2 \ln(k^2)}$ are then normally distributed.  This large number of operations is not easily vectorized and is computed sequentially.

We are able to speed up this process significantly
using instead the Box-Muller transform, which is easily vectorized. In this scheme, we draw a pair of independent numbers $A$ and $B$ from the uniform distribution over $(0,1)$ and then construct the normal distribution with $\delta x = \sqrt{-2 \ln A} \cos(2 \pi B)$ and $\delta y = \sqrt{-2 \ln A} \sin(2 \pi B)$.  We have tested that the resulting distributions of displacements are consistent with analytical PSFs to within the counting error from the finite number of timesteps, and the consistency increases as expected when the number of timesteps increases. 

To benchmark the amount of speedup gained by our scheme, we note that PSF convolution is one of several operations performed in the orbit library construction subroutine \texttt{orblib\_f.f90} in the code. This subroutine first integrates the orbits and generates the necessary reflected or rotated copies of the orbits about the symmetry axes
(see Section~\ref{sec:axi_shorttubes}).
It then computes each orbit's contribution to the 3D mass grid and projects each orbit onto the sky plane. The projected trajectories are then perturbed according to the PSF as described above.  Finally, the subroutine determines each orbit's contribution to each observed kinematic aperture on the sky and stores the associated LOSVDs.  The tasks performed in this subroutine consume
the bulk ($>90\%$) of the total runtime of the code (for one mass model); much of the remaining time is spent on performing minimizations to find optimal orbital weights.

To our surprise, our timing analysis of the various tasks executed in this subroutine (using $N_{\rm Dither} = 5$ and NGC~1453 as a test case) shows that
the PSF portion of the code (before implementing orbit axisymmetrization in Section \ref{sec:axi_shorttubes}) takes up $\sim 55\%$ of the run time, while the orbit integration itself only contributes $\sim 20\%$, and sky projections contributes the remaining $\sim 25\%$.  When we switch to the vectorized Box-Muller transform, the computation time for the PSF step becomes negligible. We are therefore able to reduce the total runtime of the code by a factor of $\sim 2$ in this test.

The speedup is even more dramatic in our axisymmetrized version when the orbits are copied azimuthally (Section~\ref{sec:axi_shorttubes}).  In this case, 80 (instead of 8) copies of each orbit are projected onto the sky and perturbed by the PSF.  We find $\sim 70\%$ of runtime is spent on the PSF portion with the original scheme,
while our new scheme reduces the runtime by a factor of $\sim 3$.

\subsection{Improve Intrinsic 3D Mass Grid}
\label{sec:massgrid}

The TriOS code uses an intrinsic 3D spatial grid to constrain the stellar component in a model to reproduce the 3D stellar density profile deprojected from the photometry of a galaxy. The code calculates the mass contributed by each orbit as it passes through a spatial bin and records this information during the stage of orbit library construction.  At the subsequent stage of orbital weight optimization, the superposition of the orbits is required to match the input mass profile within a pre-specified precision (typically 1\%) in each bin. 

In each octant of this 3D spatial grid, the code uses azimuthal and polar bins for the two angles,
each linearly spaced between 0 and $90^\circ$.  The radial bins are logarithmically spaced between $r_\mathrm{min}$ and $r_\mathrm{max}/2$, where $r_\mathrm{min}$ and $r_\mathrm{max}$ are the innermost and outermost equipotential radii used to determine the orbital energies sampled in the model. The innermost bin is then extended down to $r=0$ and the outermost is extended out to $100 r_\mathrm{max}$.  

For the outer boundary of the innermost mass bin,
we find it preferable not to base the value on $r_\mathrm{min}$, which is used for a different purpose of specifying the innermost equipotential radius for sampling orbital energy.  Instead, we modify the code to make it an independent parameter, which we set to be of similar scale as the PSF of the photometric data
since these are the data used to constrain
the deprojected 3D mass density.  To ensure that sufficient orbits are used to represent the innermost mass bins, we recommend that $r_\mathrm{min}$ be set to be smaller than the outer boundary of the innermost mass bin.  In the case of NGC~1453, we set the outer boundary of the innermost mass bin to be $0.03''$ and set $r_\mathrm{min}$ to $0.01''$. 

For similar reasons, we allow the outermost mass bin's edges to also be set independently from the outermost equipotential radius, $r_\mathrm{max}$. The remaining bin boundaries are then logarithmically sampled between the outer boundary of the innermost bin and the inner boundary of the outermost bin.

\begin{figure*}
\centering
\includegraphics[width=1.0\linewidth]{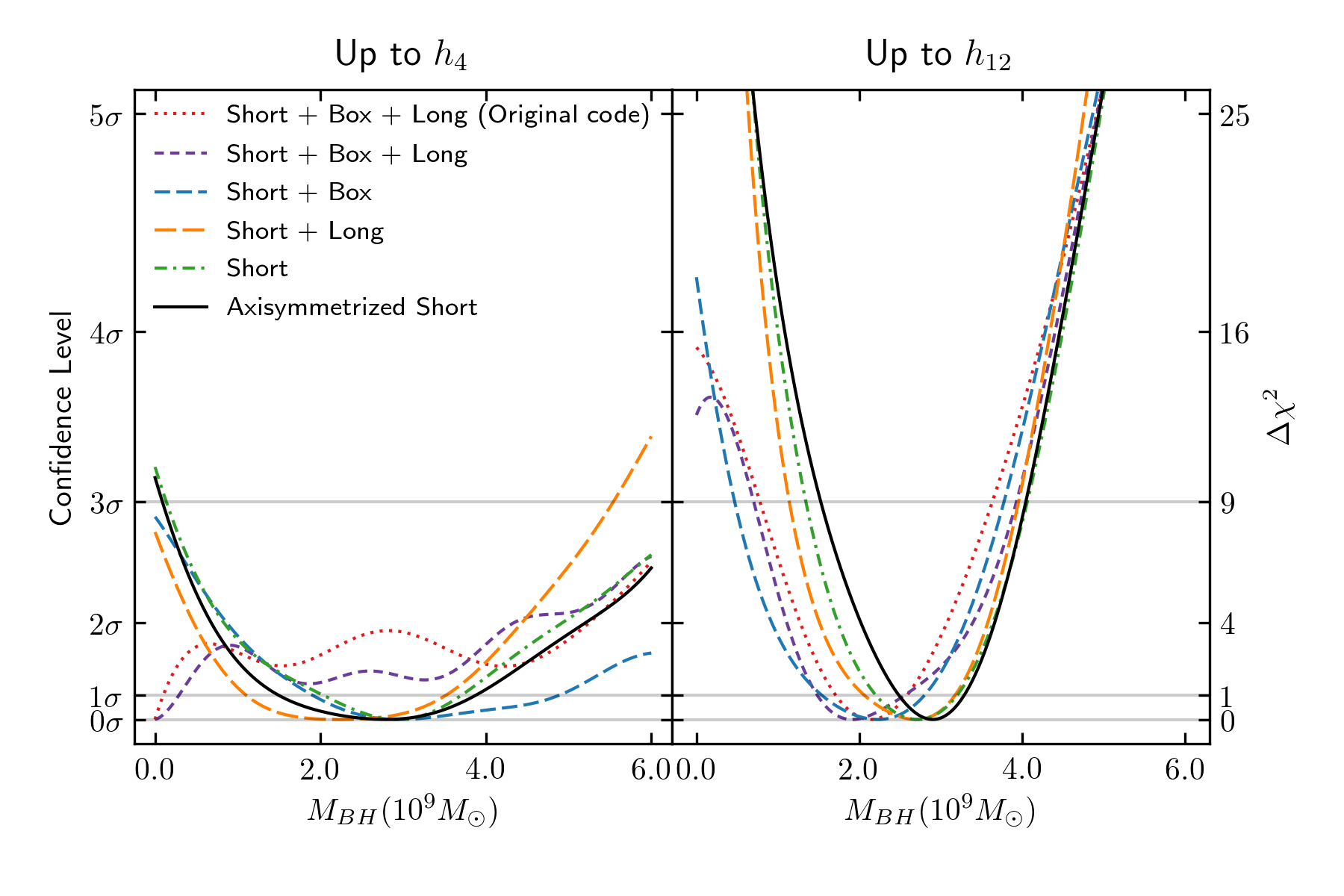}
\caption{Illustration of the changing \mbh\ constraints in NGC~1453 as the orbit model goes through the step-by-step axisymmetrization procedure described in Sections~\ref{sec:axi} and \ref{sec:minor_changes}.
The starting case (red dotted) uses the original code with typical (near) axisymmetric parameters assumed in the literature ($\psi=90.001^\circ$; see Section~\ref{sec:h4} for details). The end case (black solid) uses our final axisymmetrized code including all changes from Sections~\ref{sec:axi} and \ref{sec:minor_changes}.  The four intermediate curves have all the code fixes described in Section~\ref{sec:minor_changes}, but have different combinations of orbit types according to Sections~\ref{sec:axi_longtubes} and \ref{sec:axi_boxes}.
The left panel is for models with orbital weights chosen by fitting to the first four Gauss-Hermite moments of the LOSVDs determined from kinematic data, as is typical in the literature. The right panel uses 12 moments as constraints and shows tighter constraints on \mbh, as is reported in \citet{Liepoldetal2020}.
The 1D $\chi^2$ in \mbh\ is obtained by marginalizing over the stellar mass-to-light ratio using a smoothed 2D $\chi^2$ landscape generated by Gaussian Process regression with a squared-exponential covariance function \citep{scikit-learn}. The dark matter halo is fixed to the best-fit logarithmic halo in \citet{Liepoldetal2020}. 
}
\label{figure:versions}
\end{figure*}

\section{A Case Study: NGC 1453}
\label{sec:1453}

We use the massive elliptical galaxy NGC 1453 reported
in \citet{Liepoldetal2020} to illustrate the effects of the modifications described thus far.
In \citet{Liepoldetal2020}, we demonstrated that using more than 4 Gauss-Hermite moments was essential for obtaining robust constraints on the model LOSVDs.
Below we examine the effects in both the 4-moment and 12-moment cases, with the latter being our chosen configuration. We stress that the 4-moment case is included here only for comparison purposes since this is the typical configuration used in the literature. We have found the 4-moment case to lead to unconstrained higher moments and spurious features in the LOSVDs for NGC~1453 (Figs.~10 and 11 of \citealt{Liepoldetal2020}); the resulting $\chi^2$ in this case should therefore not be trusted. 

\begin{figure}
\centering
\includegraphics[width=\linewidth]{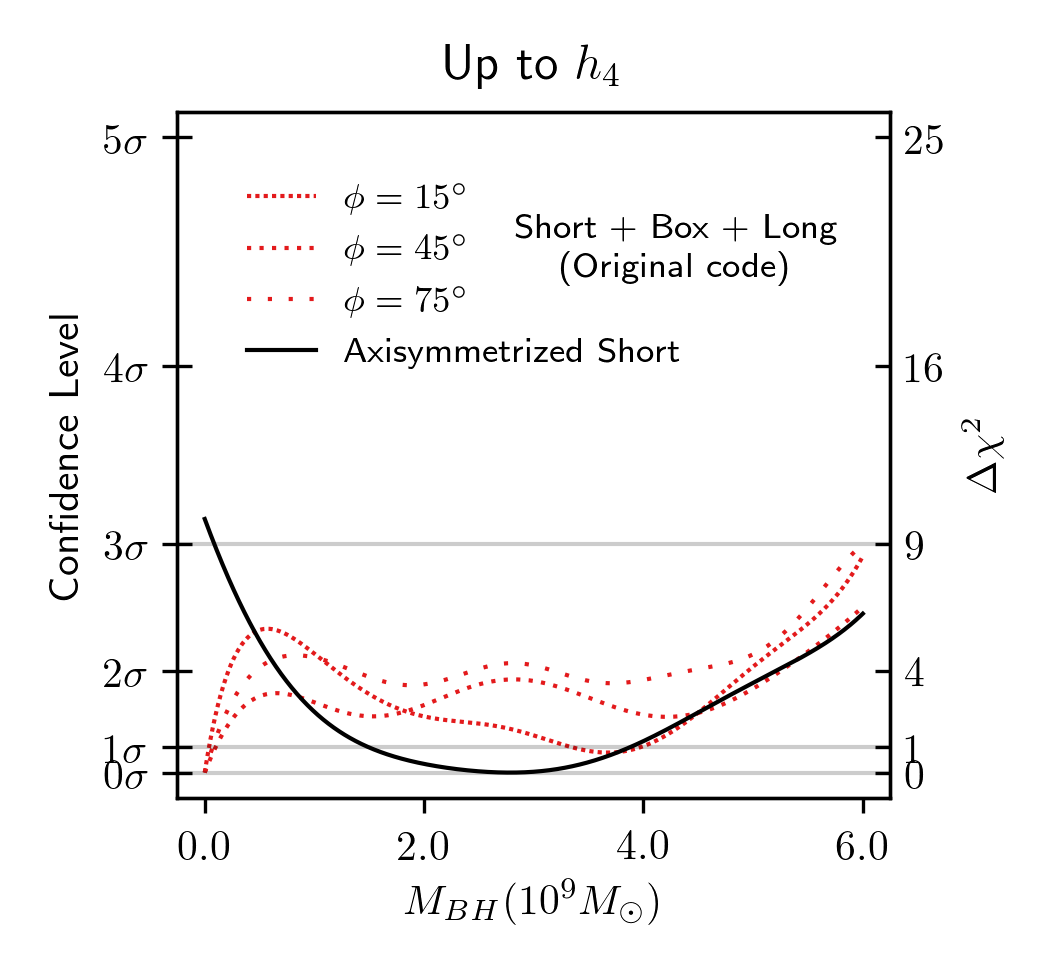}
\caption{Same as the left panel of Figure~\ref{figure:versions} but showing the azimuthal dependence of the original code when $\psi$ is chosen to be $90.001^\circ$ and all three main orbit types are included (red curves).  Our final axisymmetrized code does not depend on $\phi$ and obeys azimuthal symmetry.
  \label{figure:phi}}
\end{figure}

\subsection{Fitting up to $h_4$}
\label{sec:h4}

We begin with the case labeled ``up to $h_4$'' and ``original Leiden version'' in Figure~12 of \citet{Liepoldetal2020}.  
This case is run with the original code, $N_{\rm Dither}=3$, $N_\Theta=9$, and the viewing angles $(\theta, \phi, \psi)=(89^\circ, 45^\circ, 90.001^\circ)$, corresponding to a nearly oblate axisymmetric potential with a triaxiality parameter of $T=0.002$.  As we discussed in  Section~\ref{sec:axi_longtubes}, 
these parameters are chosen to resemble those used in earlier studies, and the models include
both the $x$-$z$ and stationary start spaces and contain all three major types of orbits: short-axis tubes, long-axis tubes and box orbits.  The left panel of Figure~\ref{figure:psi_startspace} illustrates the starting locations of both short- and long-axis tube orbits in the $x$-$z$ start space for one energy in this configuration.  

The 1D $\chi^2$ as a function of \mbh\ (marginalized over the mass-to-light ratio) is shown in the left panel of Figure~\ref{figure:versions} (red dotted curve).
As first shown in \citet{Liepoldetal2020}, the favored model in this case contains no black hole. The $\chi^2$ minimum at $\mbh=0$ here resembles the finding for the dwarf galaxy M59-UCD3 by \citet{Ahnetal2018}, which also used four Gauss-Hermite moments as constraints and a set of viewing angles with a similar deviation from axisymmetry.

Applying the code changes described in Section~\ref{sec:minor_changes} results in minor changes in the $\chi^2$ contour for NGC~1453 (purple short dashed curve in Figure~\ref{figure:versions}), but the $\mbh=0$ minimum remains.  
In the next step, we exclude the box orbits and long-axis tube orbits as described in Section~\ref{sec:axi}. The box orbits are eliminated by the simple procedure in Section~\ref{sec:axi_boxes}. To remove the long-axis tube orbits, we choose a galaxy shape that is sufficiently axisymmetric, as discussed in Section~\ref{sec:axi_longtubes}. For NGC~1453, we simply change $\psi$ from $90.001^\circ$ to $(90 + 10^{-9})^\circ$, as was done in \citet{Liepoldetal2020}. This new value is far enough from $90.0^\circ$ to avoid numerical issues in the code but is close enough to $90.0^\circ$ so that all of our orbits lie outside the long-axis tube region in the $x$-$z$ start space shown in Figure~1.

The effect of excluding these orbits on the best-fit parameter values for NGC~1453 is significant.
The preferred \mbh\ is changed from 0 to  $2.8\times10^9 M_\odot$ (green dot-dashed curve in Figure~\ref{figure:versions}a).  Before their removal, box orbits generally accounted for less than 10-35\% of total mass, while long axis tube orbits accounted for less than 2\%. 
Removing box orbits (orange long dashed curve in Figure~\ref{figure:versions}a) has a significant effect on \mbh\ because box orbits starting at different azimuthal angles are not forced to have equal weights in the original code (Section~\ref{sec:axi_boxes}).  Removing the long-axis tubes (blue dashed curve in Figure~\ref{figure:versions}a) has a significant impact likely due to their ability to fit minor-axis rotation in triaxial potentials.

In addition to excluding the box and long-axis tubes, we describe in Section~\ref{sec:axi_shorttubes} the need to enforce axisymmetry in the code by generating many azimuthally rotated copies of each short-axis tube in the $x$-$z$ start space.
For NGC~1453, we find that the main effect on the $\chi^2$ contour of this axisymmetrization procedure is to widen the minimum (black solid curve in Figure~\ref{figure:versions}a), as a broader range of orbital weights are able to fit the mass constraint for each mass model. 

The results presented thus far with the original version of the code all assumed a viewing angle of $\phi=45^\circ$. When the model galaxy is perfectly axisymmetric, this angle is irrelevant and the resulting $\chi^2$ landscape should be independent of $\phi$.
As a test, we have repeated the run with the original code (using four Gauss-Hermite moments)
with two other values of $\phi$ ($15^\circ$ and $75^\circ$) while keeping all other parameters fixed. The resulting $\chi^2$ as a function of \mbh\ for the three values of $\phi$ are shown in Figure~\ref{figure:phi}. 
The dependence on $\phi$ indicates that the mass models are indeed not consistent with axisymmetry. All three values of $\phi$ exhibit the same preference for $\mbh=0$.

\subsection{Fitting up to $h_\mathrm{12}$}

We now examine models in which the orbital weights are constrained to fit the first 12 Gauss-Hermite moments of the observed LOSVDs for NGC~1453. The first 8 moments are measured from spectroscopic observations, while the 9th-12th moments are set to 0 with an error bar based on the lower moments, as described in detail in \citet{Liepoldetal2020}. Even without any of the modifications described in this paper, \citet{Liepoldetal2020} showed that the original code performed better when 12, rather than 4, moments were used as constraints.  The right panel of Figure~12 in \citet{Liepoldetal2020} illustrated how the best-fit black hole mass moved from $\mbh=0$ for 4 moments (green curve) to $\mbh = 2.2\times 10^9 M_\odot$ for 12 moments (black curve).  The result from the original code, however, was highly dependent on the number of input moments and showed no convergence even at 12 moments.  By contrast, 
after the orbit and code modifications were implemented, the main effect of increasing the constraining kinematic moments was to tighten the error bars while leaving the best-fit values largely unchanged (left panel of Figure~12 in \citealt{Liepoldetal2020}).

Here we examine the progression of changes after each of the key modifications described in Sections~\ref{sec:axi} and \ref{sec:minor_changes} is implemented, all for the case of using 12 moments as constraints. The right panel of Figure~\ref{figure:versions} shows that
implementing the code fixes described in Section~\ref{sec:minor_changes} (purple dot-dot-dashed curve) and removing long-axis tubes (blue dot-dashed curve) move the best-fit \mbh\ by $\sim 10\%$ in comparison to $\mbh \sim 2.2\times10^9 M_\odot$ from the original code (red dotted curve).  Removing the box orbits increases \mbh\ to $\sim 2.9\times10^9 M_\odot$ (orange dot-dash-dashed and green dashed curve). The subsequent axisymmetrization of short-axis tubes (Section~\ref{sec:axi_shorttubes}) has essentially no effect (black solid curve).

To ensure that the number of orbits included in the modelling is sufficient, we tested the effect of increasing the number of orbits. We increased the density of energy sampling by a factor of 4, from 40 energy values to 160 over the same range. With 4 times the number of orbits, the best-fit \mbh\ changed by less than 3\%, and the $1\sigma$ error changed by less than 10\%, demonstrating that our results do not depend on the exact number of orbits used.

\section{Conclusion}
\label{sec:conclusion}

We have presented a revised version of the triaxial orbit superposition code by \citet{vandenBoschetal2008}, which we refer to as the TriOS code, that is capable of properly modeling axisymmetric systems.  The original code was designed for triaxial systems with (discrete) reflection symmetry along each of the three principal axes. The setup was not capable of modeling exactly axisymmetric systems in which the orbit library should respect (continuous) azimuthal symmetry about the symmetry axis.

We have implemented two main changes needed for modeling axisymmetric systems within the triaxial code: excluding all orbit types that are not allowed in an axisymmetric model, and enforcing axisymmetry among the allowed orbits. In the case of oblate axisymmetry, our recipe involves (1) axisymmetrizing the short-axis tube orbits by creating multiple copies of the orbits rotated about the symmetry axis (Section~\ref{sec:axi_shorttubes}), (2) setting the viewing angle $\psi$ to be sufficiently close to $90^\circ$ to allow no long-axis tube orbits (Section~\ref{sec:axi_longtubes}), and (3) excluding the stationary start space used to generate box orbits (Section~\ref{sec:axi_boxes}).

We have made further improvements and corrections to the code in general. We 
discussed an issue with slowly precessing quasi-planar orbits that are misclassified and are “mirrored” improperly in the orbit library (Section~\ref{sec:classification}). We also corrected a problem with the logarithmic halo implementation that prevented checking energy conservation of the integrated orbits (Section~\ref{sec:loghalo}). We achieved a factor of 2 to 3 speedup in the runtime of the code by adopting a different algorithm for modeling PSF convolution (Section~\ref{sec:psf}). Finally, we allowed the orbit sampling and mass constraints to be set independently (Section~\ref{sec:massgrid}).

For NGC~1453, 
we found the shape of the $\chi^2$ contours for \mbh\ to vary significantly as we went through  the step-by-step axisymmetrization procedure described in this paper (Figure~\ref{figure:versions}).
As we described in \citet{Liepoldetal2020},
the orbit models favored no black holes when we used the original code with typical (near) axisymmetric parameters in the literature
and four Gauss-Hermite moments to constrain the stellar LOSVDs.   
In contrast, we obtained a well constrained non-zero \mbh\ using our final axisymmetrized code including all the changes described in Sections~\ref{sec:axi} and \ref{sec:minor_changes}.

One issue that warrants further investigation in triaxial models is the equilibrium behavior of quasi-planar orbits in regions where the potential is nearly spherical, e.g., well within a SMBH's SOI, or far outside the galaxy's effective radius in a spherical dark matter halo.  
 As we discussed in Section~\ref{sec:classification}, the subset of quasi-planar orbits with precession times longer than the integration time has a nearly constant $\vec{L}$ and is misclassified and mirrored incorrectly in the original code.  Furthermore, the integration time for these orbits is not long enough to fill the allowed volume of phase space. For axisymmetric systems, we resolve these issues in our revised code described in this paper by including only short-axis tubes and enforcing axisymmetry in the orbits, while preserving $L_z$.

We also expect the severity of the orbit integration issue to vary from system to system: the better a SMBH's SOI is resolved by the available kinematic data, the more care is needed to test orbital integration time because quasi-planar orbits occupy a large fraction of the orbit library, and more orbits are deeper in the SMBH's potential and hence have longer precession times. For the NGC~1453 SMBH studied in \citet{Liepoldetal2020} and here, since our kinematic data do not reach deep inside the SOI, orbits in our mass models with precession time exceeding 200 dynamical times account for less than 4\% of luminosity within the central arcsecond.
The integration issue (and the resulting misclassification) therefore does not significantly impact our results, as is evidenced by the similarity between the solid black and green dashed curves in Figure~\ref{figure:versions}. We expect a different situation for better resolved systems such as the M87 and Milky Way SMBHs.

In future work, a straightforward solution to ensure that quasi-planar orbits are representative of their equilibrium distributions is to extend the default integration time of 200 dynamical times in the code. 
Our preliminary tests suggest that integrating the orbits up to $\sim 10$ times longer is computationally feasible, but this may still be insufficient for the orbits closest to the SMBH and in the outermost part of the galaxy where the precession times are slowest. 
A more reliable treatment of these orbits would be needed.

\acknowledgments
We thank Jonelle Walsh for useful discussions and a critical reading of the manuscript, and Karl Gebhardt for discussions about mock tests.
We also thank the anonymous referee for their helpful comments and suggestions. M.E.Q. acknowledges the support of the Natural Sciences and Engineering Research Council of Canada (NSERC), PGSD3-517040-2018. C.-P.M. acknowledges support from NSF AST-1817100, HST GO-15265, HST AR-14573, the Heising-Simons Foundation, the Miller Institute for Basic Research in Science, and the Aspen Center for Physics, which is supported by NSF grant PHY-1607611.  This work used the Extreme Science and Engineering Discovery Environment (XSEDE) at the San Diego Supercomputing Center through allocation AST180041, which is supported by NSF grant ACI-1548562.

\appendix

\section{Criterion for Existence of Long-axis Tubes}
\label{sec:ap_longaxistubes}

We use St{\"a}ckel potentials to gain insights into the existence of long-axis tubes. A potential is said to be in St{\"a}ckel form if it can be written as:

\begin{equation}
    \label{eq:stackel}
    V(\lambda,\mu,\nu) = - \frac{F(\lambda)}{(\lambda-\mu)(\lambda-\nu)} - \frac{F(\mu)}{(\mu-\nu)(\mu-\lambda)} -
    \frac{F(\nu)}{(\nu-\lambda)(\nu-\mu)},
\end{equation}
for some function $F(\tau)$ where $(\lambda,\mu,\nu)$ are ellipsoidal coordinates defined as the roots of $\tau$ in the equation
\begin{equation}
    \frac{x^2}{\tau+\alpha}+\frac{y^2}{\tau+\beta}+\frac{z^2}{\tau+\gamma}=1\,,
\end{equation}
such that $-\gamma \leq \nu \leq -\beta \leq \mu \leq -\alpha \leq \lambda$. Here, $(\alpha,\beta,\gamma)$ are constants that define the coordinate system. Such a potential is said to be separable in these coordinates. When a density corresponding to a St{\"a}ckel potential is projected in any direction to give a 2D surface density, it will have no isophotal twists~\citep{Franx1988}. Thus, we can use the viewing angles $(\theta,\phi,\psi)$ of \citet{Binney1985} to define the relationship between the primary axes of the projected and intrinsic densities. This set of viewing angles imposes a constraint on the allowed values of $(\alpha,\beta,\gamma)$ given by:

\begin{equation}
    \label{eq:foci}
    \frac{\sqrt{\beta-\alpha}}{\sqrt{\gamma-\beta}}=\sqrt{\frac{\sin^2 \theta}{\cot 2\psi \sin 2 \phi \cos \theta + \cos^2\phi (\cos^2 \theta  + 1) - 1}}.
\end{equation}
This expression follows from Equation~(B9) of \citet{Franx1988}. Orbital structure in St{\"a}ckel potentials has been well studied \citep{deZeeuw1985}. This structure is what motivated the $x$-$z$ start space described in \citet{Schwarzschild1993}. Long axis tube orbits pass through the $x$-$z$ start-space above the focal curve, defined by
\begin{equation}
    \label{eq:focal_curve}
   \frac{z^2}{\gamma-\beta} + \frac{x^2}{\alpha-\beta} =1\,.
\end{equation}
For large $x$, this curve is approximately a line given by $z \approx x \frac{\sqrt{\gamma-\beta}}{\sqrt{\beta-\alpha}}$. Therefore, the angle that this line forms with the $z$ axis can be written simply in terms of the viewing angles as
\begin{equation}
    \label{eq:eta}
 \tan\eta =\sqrt{\frac{\sin^2 \theta}{\cot 2\psi \sin 2 \phi \cos \theta + \cos^2\phi (\cos^2 \theta  + 1) - 1}} \,.
\end{equation}

Any orbits launched initially between the focal curve and the positive $z$-axis in the $x$-$z$ plane will be long axis tubes which violate axisymmetry. To effectively achieve axisymmetry, the angle $\eta$ must be small enough for no orbits to be sampled above the focal curve. Since the line defined by the angle $\eta$ is a lower bound to this curve, if all initial orbits in the positive $x$-$z$ quadrant are launched outside of the approximate angular region between the $z$-axis and the angle $\eta$, there will be no long-axis tubes in the model. 

This expression is derived for St{\"a}ckel potentials. However, in the absence of isophotal twists, we expect it to apply reasonably well to more realistic models as they can often be locally approximated by a St{\"a}ckel potential \citep{SandersBinney2015}. A central SMBH is inconsistent with a St{\"a}ckel potential and can thus destroy the ordered orbital structure. However, we suggest that Equation~(\ref{eq:eta}) could give a rough rule-of-thumb for where the boundary between long-axis and short-axis tubes will exist in models from the code, particularly at radii far from the SMBH.

\begin{figure*}
  \centering
  \includegraphics[width=\linewidth]{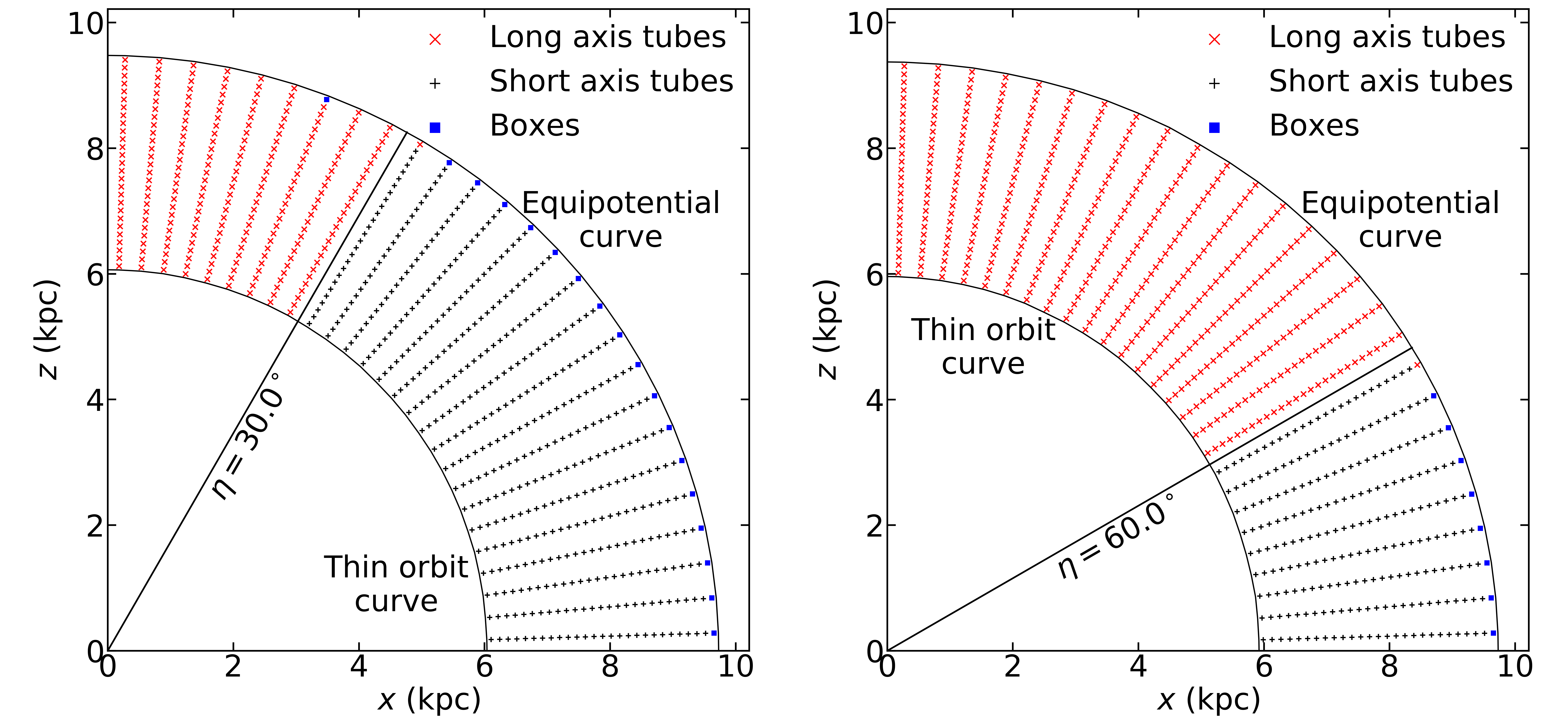}
  \caption{Same as Figure~\ref{figure:psi_startspace} but for two additional mass models with larger triaxiality: (left) triaxiality parameter $T=0.25$, (luminosity averaged) shape parameters $(u, p, q)=(0.96, 0.95, 0.77)$, and viewing angles $(\theta, \phi, \psi)=(67.62^\circ, -28.38^\circ, 86.61^\circ)$, and (right) $T=0.75$, $(u, p, q)=(0.96, 0.85, 0.79)$, and $(\theta, \phi, \psi)=(48.74^\circ, -51.33^\circ, 67.15^\circ)$.  The diagonal black line in each panel represents the angle $\eta$ given in Equation~(\ref{eta}). As in Figure~\ref{figure:psi_startspace}, this angle approximates well the boundary separating long-axis (red symbols) and short-axis (black symbols) tube orbits in the $x$-$z$ start space.  
  \label{figure:triaxial_startspace}}
\end{figure*}

The stellar mass distribution is represented by an MGE in our models. Each gaussian component is stratified on similar ellipsoids, and can thus be related to its deprojection via the equations given in \citet{Binney1985}. These equations can be rearranged to give
\begin{equation}
    \label{eq:triaxiality}
    \frac{T}{1-T}=\frac{\sin^2 \theta}{\cot 2\psi \sin 2 \phi \cos \theta + \cos^2\phi (\cos^2 \theta  + 1) - 1},
\end{equation}
where $T=(1-p^2)/(1-q^2)$ of each MGE component. For an MGE with no isophotal twists, each MGE component has the same triaxiality parameter, $T$. Thus, in this case, the angle, $\eta$, can be written simply as:
\begin{equation}
    \eta = \tan^{-1}\sqrt{\frac{T}{1 - T}} \,,
\end{equation}
where $T$ is the triaxiality parameter for each MGE component. Two examples of triaxial start spaces for NGC~1453 models are shown in Figure~\ref{figure:triaxial_startspace}. The boundary between long-axis tubes and short-axis tubes is well approximated by the angle $\eta$ for a wide range of galaxy shapes.

\section{Thin Orbit Finding}
\label{sec:ap_thinorbits}
The TriOS code uses the thin orbit curve to construct its start space. This curve has to be found numerically in the $x$-$z$ plane. For a given angle in this plane, the thin orbit radius is found by integrating test orbits starting at different radii. For each orbit, the radius of the orbit is recorded each time it passes through the $x$-$z$ plane. The thin orbit radius is found by minimizing the difference between the maximum and minimum of these radii. 

This algorithm should work for triaxial models but needs some revision in the axisymmetric case, particularly when there is no central density cusp or mass concentration. In this case, when close enough to the center, the potential should be well approximated by a harmonic oscillator. When the potential is axisymmetric, the motion can be regarded as two separate contributions: an oscillation in the $z$-axis and a closed elliptical orbit about the $z$-axis. Since the $x$-$y$ motion constitutes a closed ellipse centered on the $z$ axis, all orbits will pass through the $x$-$z$ plane at a fixed $x$ value, with some $z$ value. The orbit width is then simply set by the maximum and minimum $z$ values. Thus, for a given ray in the $x$-$z$ plane, the orbital width in this plane can be minimized by simply taking the initial radius to be as small as possible. To solve this issue when running an axisymmetric model, we instead record radii when passing through the $x$-$y$ plane. Closed ellipses will have a finite width in this plane while all thin orbits should pass through this plane in a circle of 0 width.

It is unclear how much this issue should affect the resulting orbit libraries. If orbits are sampled starting at the origin instead of the thin orbit, the result should be a less uniform sampling of angular momentum. There should also be some range of energies where the thin orbit radius is not estimated to be 0 or the correct value, but rather somewhere in between. This would result in a significantly non-uniform sampling of angular momentum since orbits passing through the $x$-$z$ plane within this radius will be undersampled relative those that do not. This issue should be essentially resolved outside of the axisymmetric limit, or if a black hole or density cusp is included. However, axisymmetric studies that use this code with no central cusp may be affected \citep{Hagenetal2019}.

\section{Mock Recovery Tests}
\label{sec:ap_mocktest}
In the above sections, we demonstrated that our changes to the TriOS code result in a consistent, non-zero SMBH mass estimate for our NGC 1453 dataset. Here, we show that the changes correctly recover the SMBH mass in a mock dataset with known parameter values.  Mock tests have been performed within various other Schwarzschild codes \citep[e.g.][]{Vallurietal2004,CrettonEmsellem2004,Magorrian2006,Siopisetal2009,VasilievValluri2020}.

For our mock galaxy, we use a flattened version of the spherical potential introduced in \citet{Siopisetal2009}. These models have an axisymmetric
gravitational potential given by
\begin{equation}
    \Phi(R,z)=\frac{1}{2}V_c^2\ln{\left(\frac{R^2+z^2/q_\Phi^2}{\mathrm{1~pc^2}}\right)}-\frac{G\mbh{}}{\sqrt{R^2+z^2}},
\end{equation}
where $q_\Phi$ is the flattening of the potential due to the extended mass distribution.
The stellar DF is 
chosen to have a Michie-like form:
\begin{equation}
    \label{eq:michie_df}
    f=A\exp{\left\{-\left[\frac{E+L_z^2/(2r_a^2)}{\sigma^2}\right]\right\}}L_z^{2N}\sqcap{(E_1,E,E_2)},
\end{equation}
where $A$ is the normalization, and $r_a$, $N$, $\sigma$, $E_1$, and $E_2$ are parameters of the model: $r_a$ is an anisotropy distance, $N$ controls the $L_z$ dependence, $\sigma$ is a characteristic velocity dispersion, and $E_1$ and $E_2$ are energy cutoffs. The symbol $\sqcap$ denotes a step function defined by
\begin{equation}
    \sqcap{(E_1,E,E_2)}=\begin{cases}
        1, & \mathrm{if}~E_1 \le E \le E_2 \\
        0, & \mathrm{otherwise}.
    \end{cases}
\end{equation}
Because $L_z$ only enters Eq.~\eqref{eq:michie_df} in even powers, there is additional freedom in how $f$ differs for positive and negative values of $L_z$. Here, we set a fixed fraction of stars to rotate in each positive direction. In this model, the stars are essentially regarded as massless tracers of the underlying potential in eq.~\ref{eq:michie_df}. Even when the potential is chosen to be spherical, the stellar distribution function can be axisymmetric. 

We use the same potential parameters as \citet{Siopisetal2009}, with $V_c=220~\mathrm{km~s^{-1}}$ and $\mbh=1.126 \times 10^8~M_\odot$. We generated two models: one model with a spherical potential ($q_\Phi=1$) to compare with \citet{Siopisetal2009}, and one model with a flattened potential ($q_\Phi=0.95$). The models both have a sphere of influence of about 10 pc. We also use the same two component DF parameters as \citet{Siopisetal2009}: the first component is a non-rotating nearly spherical bulge-like component which has $\sigma=160~\mathrm{km~s^{-1}}$, $r_a=600~\mathrm{pc}$, $N=0$, $E_1=\Phi(10~\mathrm{pc})$, $E_2=\Phi(1000~\mathrm{pc})$ with equal numbers of stars having positive and negative $L_z$; the second component is a rotating disk-like component w has $\sigma=120~\mathrm{km~s^{-1}}$, $r_a=200~\mathrm{pc}$, $N=2$, $E_1=\Phi(10~\mathrm{pc})$, $E_2=\Phi(1000~\mathrm{pc})$ with 3/4 of the stars having positive $L_z$ and 1/4 having negative $L_z$. The two components have equal numbers of stars.

We draw points in phase space from this distribution function for each star to generate mock data. We use a nearly edge-on projection, with an inclination of $\theta=89^\circ$. For the model with the flattened potential, we convolve the projected positions with a circular gaussian PSF with standard deviation 5 pc. We bin the stars into mock IFU data with a resolution of 10 pc, with a square FOV of 1000 pc. We fit an MGE to the projected surface brightness. We then run Voronoi binning on all bins with central radii $> 20~\mathrm{pc}$, resulting in 12 inner unbinned kinematic points and 108 larger outer bins. In order to keep the bins between the two models fixed, we use the Voronoi bins derived from the spherical potential. Each LOSVD is fit with a Gauss-Hermite expansion up to $h_{12}$. Gaussian noise is added to each LOSVD bin, resulting in a scatter of about 0.03 in each moment and about $0.03\sqrt{2}\sigma$
in the average velocity and velocity dispersion for each bin. We draw 20 realizations of this noise, and run the updated TriOS code for each realization. 

\begin{figure*}
\centering
\includegraphics[width=\linewidth]{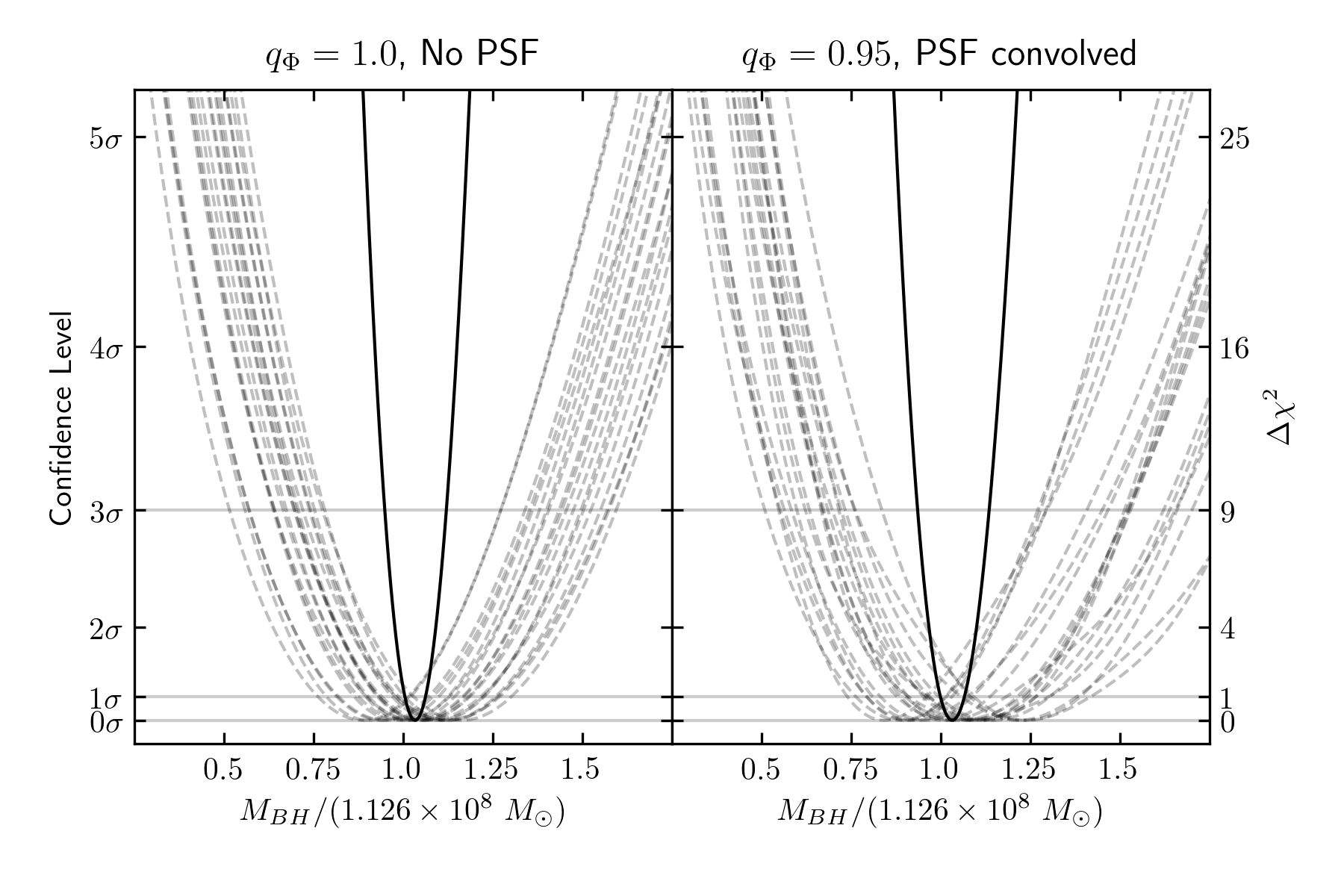}
\caption{ Illustration of the \mbh\ constraints for the mock datasets described in the text. Each dashed curve represents a separate realization of the noise, while the solid curve represents the sum over the 20 realizations. In the left panel, the potential is spherical, no PSF convolution is performed and each DF component has $5\times10^{8}$ stars. In the right panel, the model is flattened, projected stellar positions are convolved with a circular gaussian PSF with a standard deviation of 5 pc, and each DF component has $5\times10^{9}$ stars. The 1D $\chi^2$ curves are obtained by marginalizing over $V_c$ in the smoothed 2D $\chi^2$ landscape generated by Gaussian Process regression with a squared-exponential covariance function \citep{scikit-learn}. 
}
\label{figure:mock}
\end{figure*}

Figure~\ref{figure:mock} shows the resulting constraint on \mbh for each noise realization. The left panel is for the mock in the spherical potential, while the right panel is for the mock in the flattened potential. The kinematic contribution to the reduced $\chi^2$ in these realizations ranges from 0.81 to 0.94 for the spherical potential and 0.71 to 0.84 for the flattened potential, indicating a good fit to the projected kinematics for all realizations. 

In order to determine if the recovered parameter values are consistent with the input values, we calculate a combined likelihood landscape given by the product of the individual likelihoods:
\begin{equation}
    p(\mbh,V_c)=\prod_j\exp{\left(-\frac{\chi_j^2(\mbh,V_c)}{2}\right)},
\end{equation}
where $j$ indexes the 20 noise realizations~\citep{Magorrian2006}. We then calculate a combined $\chi^2$ as:
\begin{equation}
\begin{split}
    \chi^2(\mbh,V_c)&=-2\ln{p(\mbh,V_c)} \\
    &=\sum_j \chi_j^2(\mbh,V_c).
    \end{split}
\end{equation}
This is plotted as a solid line in figure~\ref{figure:mock}. The best fit SMBH mass from these combined 20 runs are $\mbh{}=(1.16^{+0.03}_{-0.03})\times10^{8}~M_\odot$ for the spherical potential, and  $\mbh{}=(1.16^{+0.03}_{-0.04})\times10^{8}~M_\odot$ for the flattened potential. In the spherical model, the recovered value of $\mbh{}$ is about 1.2 $\sigma$ from the true value. This is comparable to the results of \citet{Siopisetal2009}, where the recovered value was just within 1~$\sigma$ of true value. In the more realistic model, with a flattened potential and PSF convolution, the recovered value of $\mbh{}$ is consistent with the true value to within 1 $\sigma$. In either case, any potential systematic offset is small compared to the uncertainty on $\mbh{}$ for a single realization of about $12-14\%$.

\bibliography{axisym}

\begin{thebibliography}{}
\expandafter\ifx\csname natexlab\endcsname\relax\def\natexlab#1{#1}\fi
\providecommand{\url}[1]{\href{#1}{#1}}

\bibitem[{{Ahn} {et~al.}(2018){Ahn}, {Seth}, {Cappellari}, {Krajnovi{\'c}},
  {Strader}, {Voggel}, {Walsh}, {Bahramian}, {Baumgardt}, {Brodie},
  {Chilingarian}, {Chomiuk}, {den Brok}, {Frank}, {Hilker}, {McDermid},
  {Mieske}, {Neumayer}, {Nguyen}, {Pechetti}, {Romanowsky}, \&
  {Spitler}}]{Ahnetal2018}
{Ahn}, C.~P., {Seth}, A.~C., {Cappellari}, M., {et~al.} 2018, \apj, 858, 102

\bibitem[{Binney(1985)}]{Binney1985}
Binney, J. 1985, \mnras, 212, 767

\bibitem[{{Binney} \& {Spergel}(1984)}]{BinneySpergel1984}
{Binney}, J., \& {Spergel}, D. 1984, \mnras, 206, 159

\bibitem[{Binney \& Tremaine(2008)}]{binneytremaine2008}
Binney, J., \& Tremaine, S. 2008, Galactic Dynamics, 2nd edition), p. 75

\bibitem[{{Cappellari} {et~al.}(2006)}]{Cappellarietal2006}
{Cappellari}, M., {et~al.} 2006, \mnras, 366, 1126

\bibitem[{{Cretton} {et~al.}(1999){Cretton}, {de Zeeuw}, {van der Marel}, \&
  {Rix}}]{Crettonetal1999}
{Cretton}, N., {de Zeeuw}, P.~T., {van der Marel}, R.~P., \& {Rix}, H.-W. 1999,
  \apjs, 124, 383

\bibitem[{{Cretton} \& {Emsellem}(2004)}]{CrettonEmsellem2004}
{Cretton}, N., \& {Emsellem}, E. 2004, \mnras, 347, L31

\bibitem[{de~Zeeuw(1985)}]{deZeeuw1985}
de~Zeeuw, T. 1985, \mnras, 216, 273

\bibitem[{Feldmeier-Krause {et~al.}(2017)Feldmeier-Krause, Zhu, Neumayer,
  van~de Ven, de~Zeeuw, \& Sch??del}]{FeldmeierKrause2017}
Feldmeier-Krause, A., Zhu, L., Neumayer, N., {et~al.} 2017, \mnras, 466, 4040

\bibitem[{{Franx}(1988)}]{Franx1988}
{Franx}, M. 1988, \mnras, 231, 285

\bibitem[{{Gebhardt} {et~al.}(2000){Gebhardt}, {Richstone}, {Kormendy},
  {Lauer}, {Ajhar}, {Bender}, {Dressler}, {Faber}, {Grillmair}, {Magorrian}, \&
  {Tremaine}}]{Gebhardtetal2000a}
{Gebhardt}, K., {Richstone}, D., {Kormendy}, J., {et~al.} 2000, \aj, 119, 1157

\bibitem[{{Hagen} {et~al.}(2019){Hagen}, {Helmi}, \&
  {Breddels}}]{Hagenetal2019}
{Hagen}, J. H.~J., {Helmi}, A., \& {Breddels}, M.~A. 2019, \aap, 632, A99

\bibitem[{{Heiligman} \& {Schwarzschild}(1979)}]{HeiligmanSchwarzschild1979}
{Heiligman}, G., \& {Schwarzschild}, M. 1979, \apj, 233, 872

\bibitem[{{Hernquist}(1990)}]{Hernquist1990}
{Hernquist}, L. 1990, \apj, 356, 359

\bibitem[{{Jin} {et~al.}(2020){Jin}, {Zhu}, {Long}, {Mao}, {Wang}, \& {van de
  Ven}}]{Jinetal2020}
{Jin}, Y., {Zhu}, L., {Long}, R.~J., {et~al.} 2020, \mnras, 491, 1690

\bibitem[{{Joseph} {et~al.}(2001){Joseph}, {Merritt}, {Olling}, {Valluri},
  {Bender}, {Bower}, {Danks}, {Gull}, {Hutchings}, {Kaiser}, {Maran},
  {Weistrop}, {Woodgate}, {Malumuth}, {Nelson}, {Plait}, \&
  {Lindler}}]{Josephetal2001}
{Joseph}, C.~L., {Merritt}, D., {Olling}, R., {et~al.} 2001, \apj, 550, 668

\bibitem[{Liepold {et~al.}(2020)Liepold, Quenneville, Ma, Walsh, McConnell,
  Greene, \& Blakeslee}]{Liepoldetal2020}
Liepold, C.~M., Quenneville, M.~E., Ma, C.-P., {et~al.} 2020, \apj, 891, 4

\bibitem[{{Ma} {et~al.}(2014){Ma}, {Greene}, {McConnell}, {Janish},
  {Blakeslee}, {Thomas}, \& {Murphy}}]{Maetal2014}
{Ma}, C.-P., {Greene}, J.~E., {McConnell}, N., {et~al.} 2014, \apj, 795, 158

\bibitem[{{Magorrian}(2006)}]{Magorrian2006}
{Magorrian}, J. 2006, \mnras, 373, 425

\bibitem[{{Navarro} {et~al.}(1996){Navarro}, {Frenk}, \&
  {White}}]{Navarroetal1996}
{Navarro}, J.~F., {Frenk}, C.~S., \& {White}, S. D.~M. 1996, \apj, 462, 563

\bibitem[{Pedregosa {et~al.}(2011)Pedregosa, Varoquaux, Gramfort, Michel,
  Thirion, Grisel, Blondel, Prettenhofer, Weiss, Dubourg, Vanderplas, Passos,
  Cournapeau, Brucher, Perrot, \& Duchesnay}]{scikit-learn}
Pedregosa, F., Varoquaux, G., Gramfort, A., {et~al.} 2011, Journal of Machine
  Learning Research, 12, 2825

\bibitem[{{Pfenniger}(1984)}]{Pfenniger1984}
{Pfenniger}, D. 1984, \aap, 141, 171

\bibitem[{{Poci} {et~al.}(2019){Poci}, {McDermid}, {Zhu}, \& {van de
  Ven}}]{Pocietal2019}
{Poci}, A., {McDermid}, R.~M., {Zhu}, L., \& {van de Ven}, G. 2019, \mnras,
  487, 3776

\bibitem[{{Richstone} \& {Tremaine}(1984)}]{RichstoneTremaine1984}
{Richstone}, D.~O., \& {Tremaine}, S. 1984, \apj, 286, 27

\bibitem[{{Richstone} \& {Tremaine}(1985)}]{RichstoneTremaine1985}
---. 1985, \apj, 296, 370

\bibitem[{{Rix} {et~al.}(1997){Rix}, {de Zeeuw}, {Cretton}, {van der Marel}, \&
  {Carollo}}]{Rixetal1997}
{Rix}, H.-W., {de Zeeuw}, P.~T., {Cretton}, N., {van der Marel}, R.~P., \&
  {Carollo}, C.~M. 1997, \apj, 488, 702

\bibitem[{{Sanders} \& {Binney}(2015)}]{SandersBinney2015}
{Sanders}, J.~L., \& {Binney}, J. 2015, \mnras, 447, 2479

\bibitem[{{Schwarzschild}(1979)}]{Schwarzschild1979}
{Schwarzschild}, M. 1979, \apj, 232, 236

\bibitem[{{Schwarzschild}(1993)}]{Schwarzschild1993}
---. 1993, \apj, 409, 563

\bibitem[{Seth {et~al.}(2014)Seth, van~den Bosch, Mieske, Baumgardt, Brok,
  Strader, Neumayer, Chilingarian, Hilker, McDermid, Spitler, Brodie, Frank, \&
  Walsh}]{Sethetal2014}
Seth, A.~C., van~den Bosch, R., Mieske, S., {et~al.} 2014, \nat, 513, 398

\bibitem[{{Shapiro} {et~al.}(2006){Shapiro}, {Cappellari}, {de Zeeuw},
  {McDermid}, {Gebhardt}, {van den Bosch}, \& {Statler}}]{Shapiroetal2006}
{Shapiro}, K.~L., {Cappellari}, M., {de Zeeuw}, T., {et~al.} 2006, \mnras, 370,
  559

\bibitem[{Siopis {et~al.}(2009)Siopis, Gebhardt, Lauer, Kormendy, Pinkney,
  Richstone, Faber, Tremaine, Aller, Bender, Bower, Dressler, Filippenko,
  Green, Ho, \& Magorrian}]{Siopisetal2009}
Siopis, C., Gebhardt, K., Lauer, T.~R., {et~al.} 2009, \apj, 693, 946

\bibitem[{{Sridhar} \& {Touma}(1999)}]{SridharTouma1999a}
{Sridhar}, S., \& {Touma}, J. 1999, \mnras, 303, 483

\bibitem[{Sridhar \& Touma(2016)}]{SridharTouma2016}
Sridhar, S., \& Touma, J. 2016, \mnras, 458, 4129

\bibitem[{Thomas {et~al.}(2004)Thomas, Saglia, Bender, Thomas, Gebhardt,
  Magorrian, \& Richstone}]{Thomasetal2004}
Thomas, J., Saglia, R.~P., Bender, R., {et~al.} 2004, \mnras, 353, 391

\bibitem[{{Valluri} {et~al.}(2004){Valluri}, {Merritt}, \&
  {Emsellem}}]{Vallurietal2004}
{Valluri}, M., {Merritt}, D., \& {Emsellem}, E. 2004, \apj, 602, 66

\bibitem[{{Valluri} {et~al.}(2016){Valluri}, {Shen}, {Abbott}, \&
  {Debattista}}]{Vallurietal2016}
{Valluri}, M., {Shen}, J., {Abbott}, C., \& {Debattista}, V.~P. 2016, \apj,
  818, 141

\bibitem[{van~de Ven {et~al.}(2008)van~de Ven, De~Zeeuw, \& Van
  Den~Bosch}]{vandeVenetal2008}
van~de Ven, G., De~Zeeuw, P.~T., \& Van Den~Bosch, R. C.~E. 2008, \mnras, 385,
  614

\bibitem[{{van den Bosch} \& {de Zeeuw}(2010)}]{vandenBoschdeZeeuw2010}
{van den Bosch}, R.~C.~E., \& {de Zeeuw}, P.~T. 2010, \mnras, 401, 1770

\bibitem[{{van den Bosch} {et~al.}(2008){van den Bosch}, {van de Ven},
  {Verolme}, {Cappellari}, \& {de Zeeuw}}]{vandenBoschetal2008}
{van den Bosch}, R.~C.~E., {van de Ven}, G., {Verolme}, E.~K., {Cappellari},
  M., \& {de Zeeuw}, P.~T. 2008, \mnras, 385, 647

\bibitem[{{van der Marel} {et~al.}(1998){van der Marel}, {Cretton}, {de Zeeuw},
  \& {Rix}}]{vanderMareletal1998}
{van der Marel}, R.~P., {Cretton}, N., {de Zeeuw}, P.~T., \& {Rix}, H.-W. 1998,
  \apj, 493, 613

\bibitem[{{Vasiliev} \& {Valluri}(2020)}]{VasilievValluri2020}
{Vasiliev}, E., \& {Valluri}, M. 2020, \apj, 889, 39

\bibitem[{{Verolme} {et~al.}(2002){Verolme}, {Cappellari}, {Copin}, {van der
  Marel}, {Bacon}, {Bureau}, {Davies}, {Miller}, \& {de
  Zeeuw}}]{Verolmeetal2002}
{Verolme}, E.~K., {Cappellari}, M., {Copin}, Y., {et~al.} 2002, \mnras, 335,
  517

\bibitem[{{Walsh} {et~al.}(2012){Walsh}, {van den Bosch}, {Barth}, \&
  {Sarzi}}]{Walshetal2012}
{Walsh}, J.~L., {van den Bosch}, R.~C.~E., {Barth}, A.~J., \& {Sarzi}, M. 2012,
  \apj, 753, 79

\bibitem[{{Walsh} {et~al.}(2015){Walsh}, {van den Bosch}, {Gebhardt},
  {Yildirim}, {G{\"u}ltekin}, {Husemann}, \& {Richstone}}]{Walshetal2015}
{Walsh}, J.~L., {van den Bosch}, R.~C.~E., {Gebhardt}, K., {et~al.} 2015, \apj,
  808, 183

\bibitem[{{Walsh} {et~al.}(2017){Walsh}, {van den Bosch}, {Gebhardt},
  {Y{\i}ld{\i}r{\i}m}, {G{\"u}ltekin}, {Husemann}, \&
  {Richstone}}]{Walshetal2017}
---. 2017, \apj, 835, 208

\bibitem[{{Walsh} {et~al.}(2016){Walsh}, {van den Bosch}, {Gebhardt},
  {Y{\i}ld{\i}r{\i}m}, {Richstone}, {G{\"u}ltekin}, \&
  {Husemann}}]{Walshetal2016}
---. 2016, \apj, 817, 2

\bibitem[{{Zhu} {et~al.}(2018){Zhu}, {van de Ven}, {van den Bosch}, {Rix},
  {Lyubenova}, {Falc{\'o}n-Barroso}, {Martig}, {Mao}, {Xu}, {Jin}, {Obreja},
  {Grand }, {Dutton}, {Macci{\`o}}, {G{\'o}mez}, {Walcher},
  {Garc{\'\i}a-Benito}, {Zibetti}, \& {S{\'a}nchez}}]{Zhuetal2018b}
{Zhu}, L., {van de Ven}, G., {van den Bosch}, R., {et~al.} 2018, Nature
  Astronomy, 2, 233

\end{thebibliography}

\end{document}